\documentclass{emulateapj}
\usepackage{epsfig,natbib}
\usepackage{graphicx}
\newcommand{\ms}{\mbox{m\,s$^{-1}~$}}

\newcommand{\ks}{\mbox{km\,s$^{-1}~$}}
\newcommand{\kse}{\mbox{km\,s$^{-1}$}}
\newcommand{\mse}{\mbox{m\,s$^{-1}$}}
\newcommand{\msy}{\mbox{m\,s$^{-1}$\,yr$^{-1}~$}}
\newcommand{\msye}{\mbox{m\,s$^{-1}$\,yr$^{-1}$}}

\newcommand{\msune}{M$_{\odot}$}

\newcommand{\mjup}{M$_{\rm Jup}~$}
\newcommand{\mjupe}{M$_{\rm Jup}$}

\newcommand{\mearthe}{$M_\earth$}

\newcommand{\msinie}{$M \sin i$}
\newcommand{\vsini}{$v \sin i~$}
\newcommand{\vsinie}{$v \sin i$}

\newcommand{\mcsinie}{$M_c \sin i$}

\newcommand{\chinu}{$\chi_{\nu}$}

\newcommand{\teff}{$T_\mathrm{eff}$}
\newcommand{\feh}{\ensuremath{[\mbox{Fe}/\mbox{H}]}}
\newcommand{\rphk}{\ensuremath{R'_{\mbox{\scriptsize HK}}}}
\newcommand{\lrphk}{\ensuremath{\log{\rphk}}}
\newcommand{\caii}{\ion{Ca}{2} H \& K}
\newcommand{\caiih}{\ion{Ca}{2} H}

\shortauthors{Howard {et~al.}}
\shorttitle{Four New Giant Exoplanets}
\submitted{Submitted to ApJ}
\begin{document}
\pagenumbering{arabic}


\title{The California Planet Survey I. Four New Giant Exoplanets\altaffilmark{1}}
\author{
Andrew W.\ Howard\altaffilmark{2,3}, 
John Asher Johnson\altaffilmark{4}, 
Geoffrey W.\ Marcy\altaffilmark{2}, 
Debra A.\ Fischer\altaffilmark{5}, 
Jason T.\ Wright\altaffilmark{6}, 
David Bernat\altaffilmark{6}, 
Gregory W.\ Henry\altaffilmark{7},
Kathryn M.\,G.\ Peek\altaffilmark{2}, 
Howard Isaacson\altaffilmark{5}, 
Kevin Apps\altaffilmark{8},
Michael Endl\altaffilmark{9},
William D.\ Cochran\altaffilmark{9},
Jeff A.\ Valenti\altaffilmark{10}, 
Jay Anderson\altaffilmark{10}, and
Nikolai E.\ Piskunov\altaffilmark{11} 
}
\altaffiltext{1}{Based on observations obtained at the W.\,M.\,Keck Observatory, 
                      which is operated jointly by the University of California and the 
                      California Institute of Technology.  Keck time has been granted by both 
                      NASA and the University of California.
                      Two of the planets announced here are also based on observations 
                      obtained with the Hobby-Eberly Telescope, 
                      which is a joint project of the University of Texas at Austin, 
                      the Pennsylvania State University, Stanford University,
                      Ludwig-Maximilians-Universit\"at M\"unchen, and Georg-August-Universit\"at G\"ottingen.} 
\altaffiltext{2}{Department of Astronomy, University of California, Berkeley, CA 94720-3411 USA} 
\altaffiltext{3}{Townes Fellow, Space Sciences Laboratory, University of California, 
                        Berkeley, CA 94720-7450 USA; howard@astro.berkeley.edu}
\altaffiltext{4}{Department of Astrophysics, California Institute of Technology, MC 249-17, Pasadena, CA 91125, USA}
\altaffiltext{5}{Department of Astronomy, Yale University, New Haven, CT 06511, USA}
\altaffiltext{6}{The Pennsylvania State University, University Park, PA 16802}
\altaffiltext{7}{Center of Excellence in Information Systems, Tennessee State University, 
                        3500 John A.\ Merritt Blvd., Box 9501, Nashville, TN 37209 USA}
\altaffiltext{8}{75B Cheyne Walk, Horley, Surrey RH6 7LR, UK}
\altaffiltext{9}{McDonald Observatory, University of Texas at Austin, Austin, TX 78712, USA}
\altaffiltext{10}{Space Telescope Science Institute, 3700 San Martin Dr., Baltimore, MD 21218, USA}
\altaffiltext{11}{Department of Astronomy and Space Physics, Uppsala University, 
                        Box 515, 751 20 Uppsala, Sweden}

\begin{abstract}
We present precise Doppler measurements of four stars obtained during the past decade at Keck Observatory
by the California Planet Survey (CPS).  
These stars, namely, HD\,34445, HD\,126614, HD\,13931, and Gl\,179, 
all show evidence for a single planet in Keplerian motion.  
We also present Doppler measurements from the Hobby-Eberly Telescope (HET) for two of the stars, 
HD\,34445 and Gl\,179, that confirm the Keck detections and significantly refine the orbital parameters.
These planets add to the statistical properties of giant planets orbiting near or beyond the ice line, 
and merit follow-up by astrometry, imaging, and space-borne spectroscopy.   
Their orbital parameters span wide ranges of planetary minimum 
mass ($M$\,sin\,$i$\,=\,0.38--1.9\,\mjupe), orbital period ($P$\,=\,2.87--11.5\,yr), 
semi-major axis ($a$\,=\,2.1--5.2\,AU), and eccentricity ($e$\,=\,0.02--0.41).     
HD\,34445\,b ($P$\,=\,2.87\,yr, $M$\,sin\,$i$\,=\,0.79\,\mjupe, $e$\,=\,0.27) is a massive planet 
orbiting an old, G-type star.
We announce a planet, HD\,126614\,Ab, and an M dwarf, HD\,126614\,B, 
orbiting the metal-rich star HD\,126614 (which we now refer to as HD\,126614\,A).
The planet, HD\,126614\,Ab, has minimum mass 
$M$\,sin\,$i$\,=\,0.38\,\mjup and orbits the stellar primary with period $P$\,=\,3.41\,yr 
and orbital separation a\,=\,2.3\,AU.  
The faint M dwarf companion, HD\,126614\,B, is separated from the stellar primary by 489\,mas (33\,AU) 
and was discovered with direct observations 
using adaptive optics and the PHARO camera at Palomar Observatory.  
The stellar primary in this new system, HD\,126614\,A, 
has the highest measured metallicity (\feh\,=\,+0.56) of any known planet-bearing star.  
HD\,13931\,b ($P$\,=\,11.5\,yr, $M$\,sin\,$i$\,=\,1.88\,\mjupe, $e$\,=\,0.02) is a Jupiter analog 
orbiting a near solar twin.
Gl\,179\,b ($P$\,=\,6.3\,yr, $M$\,sin\,$i$\,=\,0.82\,\mjupe, $e$\,=\,0.21) 
is a massive planet orbiting a faint M dwarf.  
The high metallicity of Gl\,179 is consistent with the planet-metallicity correlation among 
M dwarfs, as documented recently by Johnson \& Apps. 
\end{abstract}


\section{Introduction}
\label{sec:intro}

The distributions of the masses and orbits of jovian-mass exoplanets
offer key tests of planet formation theory.  Most theories predict
that giant planets form beyond the ``snow line'' and migrate inward on
a time scale that competes with the lifetime of the protoplanetary
disk \citep{Thommes08, Ida_Lin08_v, Rice05, Alibert05, Trilling02}.
Among various theories for formation, core accretion has been shown
efficient at producing, within $\sim$3\,Myr, planets of Neptune to
Jupiter mass, orbiting within 5\,AU \citep{Benz08, Dodson-Robinson08}.

These models of giant planet formation and orbital evolution may be
directly tested against observations of giant planets found by the
Doppler method.  The models predict a clearing of gaps in the
protoplanetary disks, establishing the distribution of planet masses,
allowing a direct comparison with Doppler observed minimum masses
(\msinie).  According to the models, the disk dissipates after the
planets have undergone some inward migration, leaving them at their
current orbital distances.  The resulting predicted distribution of
semimajor axes can be compared with the observed orbits of giant
planets.  There remains potential value in both enhancing the
sophistication of planet formation theory and in observing a large
enough statistical sample of giant planets to permit robust and
informative tests of the theory.  Moreover, planet-planet interactions
among giant planets must be predicted and compared with the
distributions of orbital elements (especially eccentricity) for
systems containing multiple giant planets, e.g.\ 
\cite{Wright09,Ford_Chiang07,Ford07,Juric08}.  Giant planets also gravitationally interact
with the dust in their planetary system to shape, on timescales of
only years, the dust evolution of the planetary system 
\citep{Lisse07,Beichman07,Payne09}.

As of May 2009, 350 exoplanets have been discovered, with
remarkable properties including close-in orbits, large orbital
eccentricities, multi-planet systems, and orbital resonances
\citep{MayorUdry08, Marcy2008}.  The hot jupiters have received
the most attention, observationally and theoretically, yielding
extraordinary information about their chemical composition, internal
structure, atmospheric behavior.  However most known gas giant planets
orbit beyond 1\,AU, realizing the population that formed beyond the ice
line as predicted by theory, and as seen in our Solar System.

In 1997, we began a Doppler search for giant planets in Jovian orbits
at Keck Observatory.  We monitor over 1800 stars within 50\,pc,
with special attention given to those within 20\,pc.  We have acquired
a Doppler time baseline of well over 8\,yr for nearly all of them.
The detected long-period exoplanets reveal the distribution of their
masses, semimajor axes, and orbital eccentricities for the general
population of planetary systems.  Remarkably, the exoplanets exhibit a
sharp rise in occurrence beyond 1\,AU \citep{Johnjohn07,Cumming08}, 
indicating a great population of giant planets resides there.  
Many of these planets remain undiscovered even after 10\,yr 
of Doppler monitoring, as the amplitudes of a few meters per second 
require high Doppler precision and a clear indication of Keplerian
motion, which is challenging for orbital periods comparable to the duration
of observations.  Nonetheless, the analysis of Doppler completeness
shows that 15--18\% of all nearby stars have giant planets between 3--20\,AU
\citep{Cumming08}.  Surely, these giant planets offer strong
statistical information on the formation and subsequent dynamical
evolution of gas giants in general.

Unfortunately, with only a decade of data orbits beyond 4\,AU
are just coming into our Doppler field of view.  The recently
announced Jupiter-analog orbiting HD\,154345 with $a$\,=\,5.0\,AU (and a
circular orbit) exhibited nearly one full orbital period only after 10
full years of Doppler data were collected 
\citep{Wright_154345}.  But the number of giant planets known orbiting
beyond 1\,AU remains so small that extraordinary statistical efforts
are required to extrapolate the true underlying properties
\citep{Cumming08}.  Thus, there remains a need for enlarging the
observed population of giant planets, especially with the prospect of
follow-up observations by such instruments as 
the James Webb Space Telescope (JWST), 
the Gemini Planet Imager (GPI), and 
the Spectro-Polarimetric High-contrast Exoplanet Research (SPHERE) 
instrument for the Very Large Telescope (VLT).

In the future, knowledge of giant planets around the nearest stars will be crucial
for detecting Earth-sized planets, e.g.\ by the Space Interferometry
Mission \citep{Unwin08}, as the giant planets add a ``noise'' to the
astrometric signal \citep{Traub09}.  One concern is that
multiple giant planets orbiting beyond 2\,AU will cause curved
astrometric motion, with the linear part being (erroneously) absorbed
into the inferred proper motion.  The resulting residuals will have a
time scale of $\sim$1\,yr, constituting an astrophysical ``noise'',
compromising the detection of the terrestrial planets.  Thus, future
astrometric missions will benefit greatly from $\sim$15\,yr of radial velocity (RV) data.  
The characterization of both giant and rocky
planets will be important for future missions that image and take
spectra of planets, such as the Terrestrial Planet Finder and Darwin
\citep{Kaltenegger06, Lawson08}.  As a result we continue to survey
nearby stars that are likely targets for such surveys
\citep{Kaltenegger08}.  Here we describe Doppler measurements 
from Keck Observatory and from the Hobby-Eberly Telescope (HET) 
for four stars that shows signs of harboring a planet beyond 1\,AU.

\section{Stellar Sample and Properties}
\label{sec:stellar_props}

Among the 1800 FGKM stars we monitor at the Keck telescope for 
giant planets, 1330 of them have been monitored from 1997 to the 
present with a precision of 1--3 \mse.  The sample contains a nearly volume-limited
sample of stars within five bins of $B-V$ color, between 0.60 and
1.5. It is nearly devoid of both binary stars separated by less than 2\arcsec\ 
due to mutual contamination at the slit of the spectrometer and
magnetically active stars younger than 1\,Gyr due to their excessive
velocity ``jitter'' \citep{Wright05}.  The complete list of target stars
is given in \cite{Wright04a} including the stars that do not have
detected planets, permitting statistical analyses of the frequency of
planets, e.g. \cite{Johnjohn07b,Cumming08,Butler06}.

Only 7\% of our stars surveyed for a decade reveal clear Keplerian
Doppler variations at a 3-$\sigma$ limit of 10\,\mse.  However
\cite{Cumming08} accounted for incompleteness in our Doppler survey
and predicted that $\sim$18\% of FGK stars harbor a gas giant of
saturn mass or above within 20\,AU.  This prediction suggests that
another 11\% of our target stars harbor giant planets yet to be
revealed, with continued monitoring.  The four stars presented
here presumably harbor some of these giant planets by predicted 
\cite{Cumming08}.

\begin{deluxetable*}{lcccc}
\tablecaption{Stellar Properties}
\tablewidth{0pt}
\tablenum{1}
\label{tab:stellar_params}
\tablehead{
\colhead{Parameter}   & 
\colhead{HD\,34445} &
\colhead{HD\,126614\,A} &
\colhead{HD\,13931} &
\colhead{Gl\,179}  
}
\startdata
Spectral type & 
       G0 & 
       K0 & 
       G0 & 
       M3.5 \\  
$M_V$ & 
        4.04\,$\pm$\,0.10 & 
        4.64\,$\pm$\,0.17 & 
        4.32\,$\pm$\,0.10 & 
        11.5\,$\pm$\,0.11 \\  
$B-V$ & 
        0.661\,$\pm$\,0.015 & 
        0.810\,$\pm$\,0.004 & 
        0.642\,$\pm$\,0.015 & 
        1.590\,$\pm$\,0.015 \\  
$V$   & 
        7.31\,$\pm$\,0.03 & 
        8.81\,$\pm$\,0.002 & 
        7.61\,$\pm$\,0.03 & 
        11.96\,$\pm$\,0.03 \\  
Parallax (mas) & 
        21.5\,$\pm$\,0.7 & 
        13.8\,$\pm$\,1.0 & 
        22.6\,$\pm$\,0.7 & 
        81.4\,$\pm$\,4.0 \\  
Distance (pc) & 
        46.5\,$\pm$\,1.5 & 
        72.4\,$\pm$\,5.3 & 
        44.2\,$\pm$\,1.4 & 
        12.3\,$\pm$\,0.6 \\  
\feh & 
        $+$0.14\,$\pm$\,0.04 & 
        $+$0.56\,$\pm$\,0.04 & 
        $+$0.03\,$\pm$\,0.04 & 
        $+$0.30\,$\pm$\,0.10 \\  
$T_\mathrm{eff}$ (K) & 
        5836\,$\pm$\,44 & 
        5585\,$\pm$\,44 & 
        5829\,$\pm$\,44 & 
        3370\,$\pm$\,100 \\  
$v$\,sin\,$i$ (km\,s$^{-1}$) & 
        2.7\,$\pm$\,0.5 & 
        2.0\,$\pm$\,0.5 & 
        2.0\,$\pm$\,0.5 & 
        $<$\,1.5 \\  
log\,$g$ & 
        4.21\,$\pm$\,0.08 & 
        4.39\,$\pm$\,0.08 & 
        4.30\,$\pm$\,0.08 & 
        4.83 \\  
$L_{\star}$ ($L_{\sun}$) & 
        2.01\,$\pm$\,0.2& 
        1.21\,$\pm$\,0.19& 
        1.57\,$\pm$\,0.14& 
        0.016\,$\pm$\,0.02\\  
$M_{\star}$ ($M_{\sun}$) & 
        1.07\,$\pm$\,0.02 & 
        1.145\,$\pm$\,0.03 & 
        1.022\,$\pm$\,0.02 & 
        0.357\,$\pm$\,0.03 \\  
$R_{\star}$ ($R_{\sun}$) & 
        1.38\,$\pm$\,0.08 & 
        1.09\,$\pm$\,0.06 & 
        1.23\,$\pm$\,0.06 & 
        0.38\,$\pm$\,0.02\\  
Age (Gyr) & 
        8.5\,$\pm$\,2.0 & 
        7.2\,$\pm$\,2.0 & 
        8.4\,$\pm$\,2.0 & 
        \nodata  \\  
$S_\mathrm{HK}$ & 
        0.148 & 
        0.152 & 
        0.161 & 
        0.956 \\  
\lrphk & 
        $-$5.07 & 
        $-$5.44 & 
        $-$4.99 & 
        $-$5.20 \\  
$P_\mathrm{rot}$ (days) & 
        $\sim$\,22 & 
        $\sim$\,99 & 
        $\sim$\,26 & 
        \nodata \\  
\enddata
\end{deluxetable*}

We measure atmospheric parameters of the target stars by LTE
spectroscopic analysis of our Keck/HIRES spectra using the ``SME'' (Spectroscopy Made Easy) code
\citep{Valenti_piskunov96} as implemented by \cite{Valenti05} and
\cite{Fischer05}.  The analysis yields a best-fit estimate of \teff,
log\,$g$, \feh, \vsinie.  We could not carry out an LTE analysis of
Gl\,179 because its cool temperature resides outside the lowest
temperature for which our continuous and molecular opacities are accurate.

The luminosity of each star is determined from the apparent V-band
magnitude, the bolometric correction, and the parallax from Hipparcos.  
From \teff\ and the luminosity, we
determine the stellar mass, radius, and an age estimate by
associating those observed properties with a model from the stellar
interior calculations of \cite{Takeda07, Takeda08}.  
These properties, along with parallaxes and implied distances from 
the \cite{vanLeeuwen07} reduction of Hipparcos data,
are tabulated in Table \ref{tab:stellar_params}.

\begin{figure}
\epsscale{1.2}
\plotone{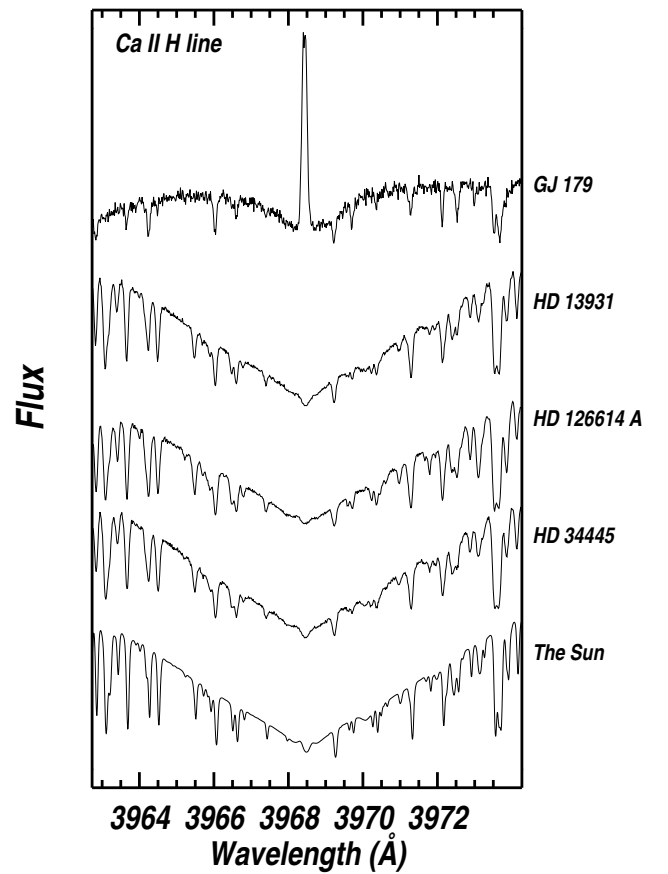}
\caption{Spectra near the \caiih\ line for all four stars discussed here. 
The emission reversals reflect magnetic activity on stars that correlates with 
photospheric velocity jitter.
The resulting chromospheric cooling rate, measured as a fraction of stellar luminosity, \lrphk, 
is listed in Table \ref{tab:stellar_params}.  
The inferred values of jitter are listed in Table \ref{tab:orbital_params}.}
\label{fig:caii}
\end{figure}

Since Gl\,179 is faint and cool, we could not use SME 
or the Takeda et al.\ stellar interior models to compute stellar properties.  
Instead, we used a variety of other techniques to estimate the properties 
listed in Table \ref{tab:stellar_params}.
We determined the mass of Gl\,179 by using the mass-luminosity calibration of
\cite{Delfosse2000}, applying its apparent K-band magnitude, $K$\,=\,6.942,
and parallax of 81.4\,mas.  
The resulting mass is $M_{\star}$\,=\,0.357\,$\pm$\,0.030\,$M_{\sun}$,
with the uncertainty from the scatter in best-fit M--L relation.  
\cite{Johnjohn09a} estimate a metallicity of \feh\,=\,0.30\,$\pm$\,0.10
for Gl\,179 based on its absolute K-band magnitude, $M_K$, and $V-K$ color.
We estimate $L_{\star}$\,=\,0.016\,$\pm$\,0.002\,$L_{\sun}$ from the bolometric correction, 
\teff\,=\,3370\,$\pm$\,100\,K from \cite{Bessel95},  
$R_{\star}$\,=\,0.38\,$\pm$\,0.02\,$R_{\sun}$ from the Stefan-Boltzmann law, 
$v$\,sin\,$i$\,$<$\,1.5\,km\,s$^{-1}$ from visual inspection of the Gl\,179 spectrum 
(Figure \ref{fig:caii}), 
and log\,$g$\,=\,4.83 from the stellar mass and radius.

We also measure the chromospheric emission in the \caii\ line cores,
providing $S_\mathrm{HK}$ values \citep{Isaacson09,Wright04a} 
on the Mt.\ Wilson system.
For each star, the time series Keck RVs and $S_\mathrm{HK}$ values are uncorrelated 
(as measured by a Pearson correlation coefficient).
We converted the average $S_\mathrm{HK}$ values for each star to
\lrphk\ as per \cite{Noyes84}, providing an estimate of the age and
rotation period of the stars.  
The \lrphk\ relation is not calibrated for cool M stars, but we include the computed value for Gl\,179 nonetheless.
All of the resulting stellar parameters are reported in Table~\ref{tab:stellar_params}.
We discuss the salient properties of each star in \S\,\ref{sec:hd34445}--\ref{sec:hip22627}.

HD\,126614 is is a binary star system composed of a bright primary, HD\,126614\,A, 
and a faint M dwarf companion, HD\,126614\,B, separated by $\sim$0\farcs5.  
The heretofore unknown companion is significantly fainter ($\Delta V$\,=\,7.8\,mag) 
and did not significantly contaminate the spectroscopic or photometric observations.  
(See \S\,\ref{sec:hd126614} for details.)

\section{Photometric Observations from Fairborn Observatory}
\label{sec:photometry}

\begin{deluxetable*}{cccccccccc}
\tabletypesize{\scriptsize}
\tablewidth{0pt}
\tablecaption{Summary of Photometric Observations from Fairborn Observatory}
\tablenum{2}
\label{tab:photometry}
\tablehead{
    \colhead{Target} & 
    \colhead{Comparison} & 
    \colhead{Comparison} & 
    \colhead{} & 
    \colhead{} & 
    \colhead{Date Range} & 
    \colhead{Target $\sigma_{\mathrm{short}}$} & 
    \colhead{Comp $\sigma_{\mathrm{short}}$} & 
    \colhead{Target $\sigma_{\mathrm{long}}$} & 
    \colhead{Comp $\sigma_{\mathrm{long}}$} \\
    \colhead{Star} & 
    \colhead{Star 1} & 
    \colhead{Star 2} & 
    \colhead{$N_{\mathrm{obs}}$} & 
    \colhead{$N_{\mathrm{yrs}}$} & 
    \colhead{(JD$ - $2,400,000)} & 
    \colhead{(mag)} & 
    \colhead{(mag)} & 
    \colhead{(mag)} & 
    \colhead{(mag)} \\
    \colhead{(1)} & 
    \colhead{(2)} & 
    \colhead{(3)} & 
    \colhead{(4)} & 
    \colhead{(5)} & 
    \colhead{(6)} & 
    \colhead{(7)} & 
    \colhead{(8)} & 
    \colhead{(9)} & 
    \colhead{(10)}
}
\startdata
 HD 34445 &  HD 34907 &  HD 35241 & 560 & 7 & 52591--54913 & 0.00155 & 0.00156 & 0.00102 & 0.00120 \\
 HD 126614 & HD 127265 & HD 124988 & 113 & 2 & 54171--54634 & 0.00158 & 0.00163 & 0.00002 & 0.00161 \\
 HD 13931 &  HD 13024 &  HD 14064 & 181 & 2 & 54730--54532 & 0.00153 & 0.00164 & 0.00158 & 0.00060 \\   
\enddata
\end{deluxetable*}

We have obtained between two and seven years of high-precision differential 
photometry of all target stars in this paper with the exception of Gl\,179.  
The observations were acquired with the T12 0.8\,m automatic 
photometric telescope (APT) at Fairborn Observatory.  This APT can 
detect short-term, low-amplitude brightness variability in solar-type 
stars due to rotational modulation in the visibility of photospheric 
starspots \citep[e.g.,][]{Henry95c} as well as longer-term variations 
associated with stellar magnetic cycles \citep{h99}.  Therefore, 
photometric observations can help to establish whether observed radial 
velocity variations in a star are due to reflex motion caused by a 
planetary companion or due to the effects of stellar activity 
\citep[e.g.,][]{Queloz2001,Paulson04}.  Photometric observations can also lead 
to the detection of planetary transits and the direct determination of 
planetary radii, as in \citet{Winn08}.

We acquired 854 nightly photometric observations of the three planetary 
host stars between 2000 November and 2009 March.  
HD\,13931, HD\,34445, and HD\,126614\,A were observed by the T12 APT, a functional duplicate of  
the T8 0.8\,m APT and two-channel photometer described in \citet{h99}.  All target stars were 
observed differentially with respect to two comparison stars in the 
Str\"omgren $b$ and $y$ bands.  One to four observations per night were 
made of each target, depending on the seasonal visibility of each star.  
To increase our photometric precision, we averaged the individual $b$ 
and $y$ differential magnitudes into a composite $(b+y)/2$ passband.  

The results of our APT observations are summarized in Table~\ref{tab:photometry}, which 
identifies the comparison stars used, the number and timespan of the 
observations, and our measures of brightness variability in the target 
and comparison stars.  All three of the observed stars have very little 
chromospheric activity (Table~\ref{tab:stellar_params} and Figure~\ref{fig:caii}).  
Therefore, given the well-known connection 
between chromospheric activity and brightness variability in solar-type 
stars \citep[see e.g.,][Fig. 11]{h99}, we did not expect to find significant 
photometric variability in these four targets.

We use the standard deviation as a simple metric for brightness variability 
of our target and comparison stars, and we measure the variability on 
both night-to-night ($\sigma_{\mathrm{short}}$) and year-to-year ($\sigma_{\mathrm{long}}$) 
timescales.  In general, the standard deviations of individual measurements 
of pairs of constant stars fall in the range 0.0012--0.0017 mag for 
these telescopes.  Column~7 of Table~\ref{tab:photometry} gives the standard deviation of 
the individual target minus comparison star differential magnitudes 
averaged across both comparison stars and across all observing seasons.  
Column~8 gives the standard deviation of the individual differential 
magnitudes between the two comparison stars averaged across all observing 
seasons.  We compute the ($\sigma_{\mathrm{long}}$) values in the same manner but 
with the yearly mean differential magnitudes rather than individual 
nightly magnitudes.  The photometric results for each of the four stars 
are briefly discussed in the following sections.  Given the moderately 
long orbital periods and the current uncertainties in the orbital 
parameters (Table~3), a search for planetary transits in any of the four  
target stars is impractical.

\section{Spectroscopic Observations and Keplerian Orbital Solutions}
\label{sec:orbital}

\subsection{Keck Observations and Doppler Reduction}

We observed HD\,34445, HD\,126614\,A, HD\,13931, and Gl\,179 
using the HIRES echelle spectrometer \citep{Vogt94} on the 10-m Keck I telescope.  
These stars were each observed 30--70 times each over 10--12\,yr.  
All observations were made with an iodine cell mounted directly in front of the 
spectrometer entrance slit.  The dense set of molecular absorption lines imprinted 
on the stellar spectra provide a robust wavelength fiducial 
against which Doppler shifts are measured, 
as well as strong constraints on the shape of the spectrometer instrumental profile at 
the time of each observation \citep{Marcy92,Valenti95}.

We measured the Doppler shift from each star-times-iodine spectrum using a 
modeling procedure modified from the method described by \citet{Butler96b}.  
The most significant modification is the way we model the
intrinsic stellar spectrum, which serves as a reference point for the relative Doppler
shift measurements for each observation.  Butler et al.\ use a version of the
\citet{Jansson95} deconvolution algorithm to remove the spectrometer's instrumental profile
from an iodine-free template spectrum.  We instead use a new deconvolution
algorithm developed by one of us (J.\,A.\,J.) that employs a more effective regularization
scheme, which results in significantly less noise amplification and improved
Doppler precision.  

\begin{figure}
\epsscale{1.2}
\plotone{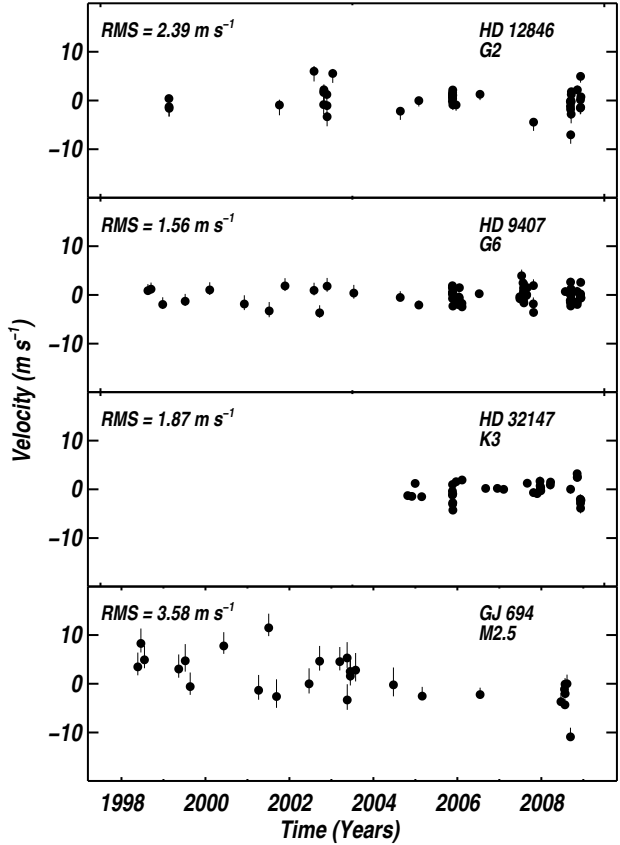}
\caption{Radial velocity time series for four stable stars in our Keck Doppler survey.
These stars demonstrate long-term velocity stability over a wide range of spectral types.
Gl\,694 (M2.5, V=10.5) shows that our velocity precision is reduced around 
faint, late spectral type stars, but that it is sufficient to detect the planet around Gl\,179 (M3.5, V=11.96).
The binned velocities with measurement uncertainties (but not jitter) are plotted.  
Panels are labeled with star name, spectral type, and rms to a linear fit.}
\label{fig:standard_stars}
\end{figure}

Figure \ref{fig:standard_stars} shows the radial velocity time series for four stable stars with
characteristics similar to the planet-bearing stars presented herein, 
demonstrating our measurement precision over the past decade on a variety of spectral types. 
In 2004 August, the Keck HIRES spectrometer was upgraded with a new detector. 
The previous 2K\,$\times$\,2K pixel Tektronix CCD was replaced by an array of 
three 4K\,$\times$\,2K pixel MIT-LL CCDs. 
The new detector produces significantly higher velocity precision due to its improved 
charge transfer efficiency and charge diffusion characteristics, smaller pixels  
(15\,$\mu$m vs.\ 24\,$\mu$m), higher quantum efficiency, 
increased spectral coverage, and lower read noise.
Our post-upgrade measurements exhibit a typical long-term rms scatter of $\sim$1.5\,\ms 
for bright, quiescent stars, compared to $\sim$2.5\,\ms for pre-upgrade measurements.  
(Measurements prior to JD 2,453,237 are pre-upgrade.)
The pre- and post-upgrade measurements also lack a common velocity zero point, 
but we fit for and corrected this offset to within $\sim$2\,\ms 
for every star observed at Keck using a large set of 
stable stars with many pre- and post-upgrade observations.
To further limit the impact of the velocity discontinuity, 
we let the offset float in the Keplerian fits below, 
effectively treating pre-upgrade and post-upgrade observations as coming 
from two different telescopes 
(except for HD\,13931, which does not have enough post-upgrade measurements 
to allow for a floating offset).  

With the exception of Gl\,179, these planet-bearing stars are bright G and K stars, 
and the Keck RVs have median single measurement uncertainties of 1.2--1.4\,\mse.  
In some instances we made two or more consecutive observations of the same star and 
binned the velocities in 2\,hr time bins, thereby reducing the associated measurement uncertainties.   
Gl\,179 is substantially fainter ($V$\,=\,11.96), and its median measurement uncertainty 
of 4.2\,\ms is dominated by Poisson noise from photon statistics.
For each of these stars, the measurement uncertainty is the weighted standard deviation  
of the mean of the velocity measured from each of the $\sim$700 2\,$\mathrm{\AA}$ chunks 
in the echelle spectrum \citep{Butler96b}.  

\subsection{HET Observations and Doppler Reduction}

We observed two of the stars, HD\,34445 and Gl\,179, with the Hobby-Eberly Telescope (HET) 
\citep{Ramsey98} at McDonald Observatory, 
as part of the on-going Doppler surveys of FGK-type stars \citep{Cochran04} 
and M dwarfs \citep{Endl03}. 
We use the High-Resolution-Spectrograph (HRS) (Tull~1998) 
in combination with an I$_2$-cell at a resolving
power of $R=60,000$ to measure precise radial velocities. 
The observing strategy and data reduction pipeline are detailed in \citet{Cochran04} 
and a description of our I$_2$-data modeling code 
``Austral'' can be found in \citet{Endl00}.

HET observations of HD~34445 begun in 2003 October and 
we collected a total of 45 HRS spectra since then. 
We started RV monitoring of Gl~179 a year later, in 2004 October, 
and so far we have obtained 11 spectra of this faint M dwarf.

\begin{deluxetable*}{lcccc}
\tablecaption{Single Planet Keplerian Orbital Solutions}
\tablenum{3}
\label{tab:orbital_params}
\tablewidth{0pt}
\tablehead{
\colhead{Parameter}   & 
\colhead{HD\,34445} &
\colhead{HD\,126614\,A} &
\colhead{HD\,13931} &
\colhead{Gl\,179}  
}
\startdata
$P$ (yr) & 
        2.87\,$\pm$\,0.03 & 
        3.41\,$\pm$\,0.05 & 
        11.5\,$\pm$\,1.1 & 
        6.26\,$\pm$\,0.16 \\  
$P$ (d) & 
        1049\,$\pm$\,11 & 
        1244\,$\pm$\,17 & 
        4218\,$\pm$\,388 & 
        2288\,$\pm$\,59 \\  
$e$ & 
        0.27\,$\pm$\,0.07 & 
        0.41\,$\pm$\,0.10 & 
        0.02\,$\pm$\,0.05 & 
        0.21\,$\pm$\,0.08 \\  
$K$ (m\,s$^{-1}$) &  
        15.7\,$\pm$\,1.4 & 
        7.3\,$\pm$\,0.7 & 
        23.3\,$\pm$\,1.4 & 
        25.8\,$\pm$\,2.2 \\  
$T_p$ (JD -- 2,440,000) & 
        13,781\,$\pm$\,48 & 
        13,808\,$\pm$\,52 & 
        14,494\,$\pm$\,904 & 
        15,140\,$\pm$\,104 \\  
$\omega$ (deg) & 
        104\,$\pm$\,19 & 
        243\,$\pm$\,19 & 
        290\,$\pm$\,78 & 
        153\,$\pm$\,24 \\  
$dv/dt$ (\msye) & 
        $\equiv$\,0.0& 
        16.2\,$\pm$\,0.2& 
        $\equiv$\,0.0 & 
        $\equiv$\,0.0 \\  
$M$\,sin\,$i$ (\mjupe) & 
        0.79\,$\pm$\,0.07 & 
        0.38\,$\pm$\,0.04 & 
        1.88\,$\pm$\,0.15 & 
        0.82\,$\pm$\,0.07 \\  
$a$ (AU) & 
        2.07\,$\pm$\,0.02 & 
        2.35\,$\pm$\,0.02 & 
        5.15\,$\pm$\,0.29 & 
        2.41\,$\pm$\,0.04 \\  
$N_\mathrm{obs}$ (Keck, binned) & 
        68 & 
        70 & 
        39 & 
        30 \\  
$N_\mathrm{obs}$ (HET, binned) & 
        50 & 
        \nodata & 
        \nodata & 
        14 \\  
Median binned uncertainty (Keck, \mse) & 
        1.4 & 
        1.3 & 
        1.3 & 
        4.7 \\  
Median binned uncertainty (HET, \mse) & 
        2.7 & 
        \nodata & 
        \nodata & 
        7.2 \\  
Assumed Keck pre-upgrade jitter (\mse) & 
       3.0 & 
       3.0 & 
       3.0 & 
       3.0 \\  
Assumed Keck post-upgrade jitter (\mse) & 
       2.0 & 
       2.0 & 
       2.0 & 
       2.5 \\  
Assumed HET jitter (\mse) & 
        2.0 & 
        \nodata & 
        \nodata & 
        2.5 \\  
rms to fit (\mse) & 
        7.31 & 
        3.99 & 
        3.31 & 
        9.51 \\  
$\sqrt{\chi^2_\nu}$ & 
        2.39 & 
        1.42 & 
        1.01 & 
        1.78 \\  
\enddata
\end{deluxetable*}

\subsection{Keplerian Models}
\label{sec:keplerian_models}

For each of the four stars below, 
we present a single-planet Keplerian fit to the measured velocities.  
For HD\,126614\,A, we add a linear velocity trend to the model.
Table \ref{tab:orbital_params} summarizes the observing statistics, best-fit Keplerian orbital parameters, 
and measures of the goodness-of-fit for these planets.
In this subsection, we describe the Keplerian modeling and related topics associated for all of the stars.  
We discuss the planet-bearing stars individually in \S\,\ref{sec:hd34445}--\ref{sec:hip22627}.

With the measured velocities for each star we performed a thorough search  
of the orbital parameter space for 
the best-fit, single-planet Keplerian orbital model using the partially-linearized, 
least-squares fitting procedure described in \citet{Wright09b}.  
Each velocity measurement was assigned a weight, $w$,
constructed from the quadrature sum of the measurement uncertainty 
($\sigma_{\mathrm{RV}}$) and a jitter term ($\sigma_{\mathrm{jitter}}$), 
i.e.\ $w$ = 1/($\sigma_{\mathrm{RV}}^2+\sigma_{\mathrm{jitter}}^2$).  
Following \citet{Wright05}, and based on the values of $S_\mathrm{HK}$, $M_V$, and $B-V$, 
we estimate $\sigma_{\mathrm{jitter}}$ for each of the stars (Table \ref{tab:orbital_params}).
These empirical estimates account for RV variability due to myriad sources of stellar noise, 
including the velocity fields of magnetically-controlled photospheric turbulence 
(granulation, super-granulation, and meso-granulation),
rotational modulation of stellar surface features, stellar pulsation, and magnetic cycles, 
as well as undetected planets 
and uncorrected systematic errors in the velocity reduction \citep{Saar98,Wright05}. 
We adopt larger values of jitter for pre-upgrade Keck measurements 
($\sigma_{\mathrm{jitter}}$\,=\,3\,\mse) than for HET measurements and 
post-upgrade Keck measurements ($\sigma_{\mathrm{jitter}}$\,=\,2--2.5\,\mse).  
The difference accounts for the slightly larger systematic errors incurred 
in the reduction of pre-upgrade spectra.  

The Keplerian parameter uncertainties for each planet were derived using a Monte Carlo 
method \citep{Marcy05} and do not account for correlations between parameter errors.  

For stars with small Doppler amplitudes (HD\,34445 and HD\,126614\,A), 
we also explicitly considered the null hypothesis---that the velocity periodicity 
represented by the Keplerian orbital fit arose from chance fluctuations in the velocities---by 
calculating a false alarm probability (FAP) based on $\Delta\chi^2$, 
the goodness of fit statistic \citep{Howard09a,Marcy05,Cumming08}.
These FAPs compare the measured data to 
1000 scrambled data sets drawn randomly with replacement from the original measurements.  
(The linear velocity trends seen in HD\,126614\,A was subtracted from the velocities before scrambling.)
For each data set we compare a best-fit Keplerian model to 
the null hypothesis (a linear fit to the data) by computing   
$\Delta\chi^2$\,=\,$\chi^2_{\mathrm{lin}}$\,$-$\,$\chi^2_{\mathrm{Kep}}$, 
where $\chi^2_{\mathrm{lin}}$ and $\chi^2_{\mathrm{Kep}}$ 
are the values of $\chi^2$ for linear and Keplerian fits to the data, respectively.   
The $\Delta\chi^2$ statistic measures the improvement in the fit of a Keplerian model 
compared to a linear model of the same data.  
The FAP is the fraction of scrambled data sets that have a larger value of $\Delta\chi^2$ 
than for the unscrambled data set.  
That is, the FAP measures the fraction of scrambled data sets where the statistical improvement from a 
best-fit Keplerian model over a linear model is greater than the statistical improvement of 
a Keplerian model over a linear model for the actual measured velocities.
We use $\Delta\chi^2$ as the goodness-of-fit statistic, 
instead of other measures such as $\chi_{\nu}$ for a Keplerian fit, 
to account for the fact that the scrambled data sets, 
drawn from the original velocities \textit{with replacement}, 
have different variances, which sometimes artificially improve the fit quality 
(i.e.\  some scrambled data sets contain fewer outlier velocities and have lower rms).  
It is important to note that this FAP does not measure the probability of 
non-planetary sources of true velocity variation masquerading as a planetary signature.

\section{HD\,34445}
\label{sec:hd34445}

\subsection{Stellar Characteristics}

HD\,34445 (=\,HIP\,24681) is spectral type G0, with $V$\,=\,7.31, $B$\,--\,$V$\,=\,0.66, and 
$M_V$\,=\,4.04, placing it $\sim$0.8 mag above the main sequence
as defined by \citet{Wright05}.   
It is chromospherically quiet with $S_\mathrm{HK}$\,=\,0.15 and \lrphk\,=\,$-$5.07 \citep{Isaacson09}, 
implying a rotation period of $\sim$22 days \citep{Noyes84}.
\cite{Valenti05} measured a super-solar metallicity of \feh\,=\,+0.14 using SME.  
Its placement above the main sequence and low chromospheric activity are 
consistent with an old, quiescent star of age 8.5\,$\pm$\,2\,Gyr.

\subsection{Photometry from Fairborn Observatory}

We have 560 individual photometric observations of HD\,34445 spanning seven 
consecutive observing seasons.  The mean short-term standard deviation of 
HD\,34445 is comparable to the mean standard deviations of the comparison 
stars (Table~\ref{tab:photometry}; columns 7 \& 8); both are within the range of our typical 
measurement precision for single observations.  We performed periodogram 
analyses over the range of 1--100 days on the seven individual observing 
seasons and on the entire data sets and found no significant periodicity 
in either HD\,34445 or the comparison stars.  Therefore, our short-term 
variability measurements of 0.00155 and 0.00156 mag are upper limits to 
any real nightly brightness variability in the target and comparison stars.

Similarly, the long-term standard deviations of HD\,34445 and the 
comparison stars are similar (Table~\ref{tab:photometry}, columns 9 \& 10), indicating 
that we have not resolved any intrinsic year-to-year variability in 
HD\,34445.  \citet{h99} and \citet{Wright_154345} present data sets that have
seasonal means with standard deviations as low as 0.0002 mag, which we take
as our mesurement precision for seasonal mean magnitudes.  Thus, our 
non-detection of long-term variability in HD\,34445 may be compromised 
by low-level variability in one or both of the comparison stars.

\subsection{Doppler Observations and Keplerian Fit}

\begin{figure}
\epsscale{1.2}
\plotone{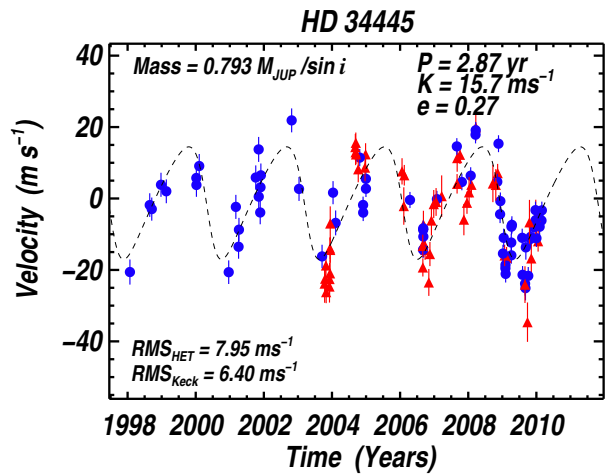}
\caption{Measured velocity vs.\ time for HD\,34445
    with the associated best-fit Keplerian model (\textit{dashed line}).
    Filled blue circles represent measurements from Keck, 
    while filled red triangles represent HET measurements.
    The error bars show the quadrature sum of measurement uncertainties and jitter.}
\label{fig:34445_ts_1p_1}
\end{figure}

We monitored the radial velocity of HD\,34445 at Keck for the past 12\,yr
and at the HET for the past 6\,yr.
Table \ref{tab:rv_34445} gives the Julian Dates of the observations, 
the measured (relative) radial velocities, the associated measurement uncertainties (excluding jitter), 
and the telescope used.  
The 118 velocities listed in Table \ref{tab:rv_34445} have an rms of 12.0\,\mse 
and are plotted as a time series in Figure \ref{fig:34445_ts_1p_1} 
along with the best-fit, single-planet Keplerian model.

The velocities plotted in Figure \ref{fig:34445_ts_1p_1} reveal a $\sim$2.8\,yr periodicity 
that is apparent by visual inspection.  
We searched periods near 2.8\,yr, and a wide variety of other periods, 
to find a single-planet Keplerian model with best-fit parameters, 
$P$\,=\,2.87\,$\pm$\,0.03\,yr, 
$e$\,=\,0.27\,$\pm$\,0.07, 
and $K$\,=\,15.7\,$\pm$\,1.4\,\mse,
implying a planet of minimum mass $M$\,sin\,$i$\,=\,0.79\,\mjup 
orbiting with semimajor axis $a$\,=\,2.07\,AU. 
The full set of orbital parameters is listed in Table \ref{tab:orbital_params}.\footnote{Advanced 
mention of the existence of this planet for HD\,34445 was made in \cite{Fischer05}.}
This joint fit has an rms of 7.31\,\ms with \chinu\,=\,2.39.  
The planetary signal was detected in the Keck and HET data sets individually, 
which have an rms of 6.39\,\ms and 7.95\,\ms about the joint fit, respectively.

We also considered the null hypothesis---that the observed RV measurements are the chance 
arrangement of random velocities masquerading as a coherent signal---by calculating an FAP.  
As described in \S\,\ref{sec:keplerian_models}, 
we computed the improvement in $\Delta\chi^2$ from 
a constant velocity model to a Keplerian model (without a trend) for 10$^3$ scrambled data sets.  
We found that no scrambled data set had a larger value for $\Delta\chi^2$ 
than for the measured velocities, implying an FAP of less than $\sim$0.001 for this scenario.

While the single-planet model appears secure, the rms of the velocity residuals (6--8\,\mse) 
is a factor of $\sim$2 higher than expected based on the measurement uncertainties 
and jitter.  Two possible explanations for this excess variability are underestimated jitter 
and additional planets.
Jitter seems an unlikely explanation given this star's metallicity, color, and modest evolution.  
As a comparison, the 5 stars most similar to HD\,34445 
($B$\,--\,$V$ and $M_V$ within 0.05\,mag) 
with 10 or more Keck observations have velocity rms in the range 2.6--4.9\,\mse.  
We searched a wide range of possible periods for a second planet
and found several candidates, the strongest of which has $P_c$\,=\,117\,d,  \mcsinie\,=\,52\,\mearthe, 
and an FAP of a few percent.
A significant number of additional measurements are required to confirm this planet and rule out other periods.

\section{HD\,126614\,A}
\label{sec:hd126614}

\subsection{Stellar Characteristics}

HD\,126614 (=\,HIP\,70623) is identified in the Henry Draper and Hipparcos catalogs as a single star.  
As described in \S\,\ref{sec:hd126614_ao}, we directly detected by adaptive optics (AO) a faint M dwarf companion 
separated by 489\,mas from the bright primary.
We refer to the bright primary star as HD\,126614\,A and the faint companion as HD\,126614\,B.  
These stars are unresolved in the Doppler and photometric variability observations described below.  
The planet announced below orbits HD\,126614\,A and is named HD\,126614\,Ab.  
In addition, HD\,126614\,A is orbited by a second M dwarf, NLTT\,37349, 
in a much wider orbit \citep{Gould04}.
This outer stellar companion is separated from HD\,126614\,A by 42\arcsec\ and does not 
contaminate the Doppler or photometric observations.  

HD\,126614\,A is spectral type K0, 
with $V$\,=\,8.81, $B$\,--\,$V$\,=\,0.81, and 
$M_V$\,=\,4.64, placing it $\sim$1.2 mag above the main sequence.   
It is chromospherically quiet with $S_\mathrm{HK}$\,=\,0.15 and \lrphk\,=\,$-$5.44 \citep{Isaacson09}, 
implying a rotation period of 99\,d, 
which is off the scale of the calibration, but suggests that it is longer than 50\,d \citep{Noyes84}.
\cite{Valenti05} measured an extremely high metallicity of \feh\,=\,+0.56 using SME.  
This is the highest metallicity measured in the 1040 stars in the SPOCS catalog \citep{Valenti05}.  
The low chromospheric activity ($S_\mathrm{HK}$\,=\,0.15) is consistent with 
the age estimate of 7.2\,$\pm$\,2.0\,Gyr from the SME analysis \citep{Valenti05}.

\subsection{Photometry from Fairborn Observatory}

We have only 113 photometric observations of HD\,126614\,A covering two observing seasons.
Periodogram analyses over the range of 1--100 days found no significant 
periodicity in either HD\,126614\,A or its comparison stars.  The short-term 
variability of 0.00158 mag in the target star is comparable to 0.00163 mag 
in the comparison stars and to our measurement precision.  The measurement 
of long-term variability in HD\,126614\,A is compromised by low-amplitude 
variability in one or both comparison stars.  Thus, 0.0016 mag serves as 
an upper limit to both long- and short-term variability in HD\,126624\,A.

\subsection{Doppler Observations and Keplerian Fit}
\label{sec:126614_RV}

\begin{figure}
\epsscale{1.2}
\plotone{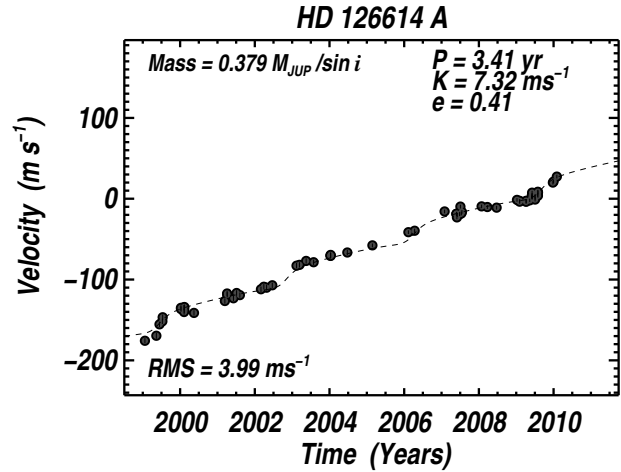}
\caption{Measured velocity vs.\ time for HD\,126614\,A (\textit{filled circles})
    and best-fit Keplerian model (\textit{dashed line}).
    The strong linear trend of $dv/dt$\,=\,16.2\,\msy is clearly seen.  
    The error bars that represent the quadrature sum of measurement uncertainties and jitter 
    are not visible on this scale.
    These data are plotted again with the linear trend removed in Figure \ref{fig:126614_ts_1p_1_notrend}.}
\label{fig:126614_ts_1p_1}
\end{figure}

\begin{figure}
\epsscale{1.2}
\plotone{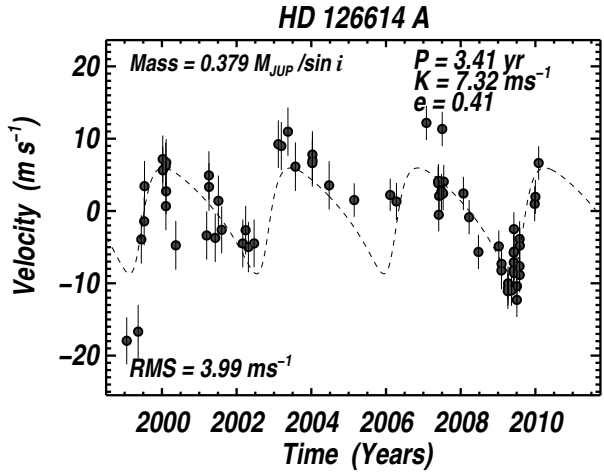}
\caption{The same measured velocities (\textit{filled circles}) and 
    single planet Keplerian model (\textit{dashed line}) for HD\,126614\,A as plotted in 
    Fig.\ \ref{fig:126614_ts_1p_1}, except the linear trend  was subtracted before plotting.}
\label{fig:126614_ts_1p_1_notrend}
\end{figure}

We monitored the radial velocity of HD\,126614\,A at Keck for the past 11\,yr.
Table \ref{tab:rv_126614} gives the Julian Dates of the observations, 
the measured (relative) radial velocities, and the associated measurement uncertainties (excluding jitter).  
The 70 velocities listed in Table \ref{tab:rv_126614} 
and plotted as a time series in Figure \ref{fig:126614_ts_1p_1}
display a strong linear trend of 16.2\,\msy with an rms of 6.6\,\ms about that trend.  
For clarity, the same velocities are plotted in Figure \ref{fig:126614_ts_1p_1_notrend} 
with the trend subtracted.

After subtracting the linear trend, 
a Lomb--Scargle periodogram of the HD\,126614\,A velocities 
reveals a strong periodicity near 3.4\,yr.
This periodicity closely matches the period of the best-fit, single-planet Keplerian model, 
which was found after a thorough search of the Keplerian parameters.  
Our model of a single planet plus a linear velocity trend 
has a distinctive staircase-like appearance (Figure \ref{fig:126614_ts_1p_1}) and 
the following best-fit parameters:   
$P$\,=\,3.41\,$\pm$\,0.05\,yr, 
$e$\,=\,0.41\,$\pm$\,0.10, 
$K$\,=\,7.3\,$\pm$\,0.7\,\mse,
and $dv/dt$\,=\,16.2\,$\pm$\,0.2\,\msye, 
implying a planet of minimum mass $M$\,sin\,$i$\,=\,0.38\,\mjup 
orbiting with semimajor axis $a$\,=\,2.35\,AU.  
The full set of orbital parameters is listed in Table \ref{tab:orbital_params}. 
The time series velocities (with trend subtracted) are plotted with the Keplerian model 
in Figure \ref{fig:126614_ts_1p_1_notrend}.
 
This model of a single planet plus linear trend has rms velocity residuals of 3.99\,\ms and $\chi_{\nu} = 1.42$,
implying an adequate fit.  
We computed an FAP for the single-planet model using the $\Delta\chi^2$ statistic, 
as described in \S\,\ref{sec:keplerian_models}.  
We found that no scrambled data set had a larger value for $\Delta\chi^2$ 
than for the measured velocities, implying an FAP of less than $\sim$0.001 for this scenario.

\subsection{Source of Linear RV Trend}

The significant linear trend of 16.2\,\msy is likely due to a long-period stellar or planetary companion.
We tried fitting the observed velocities with a 2-planet model, 
using the Keplerian parameters of HD\,126614\,Ab and an outer planet with a wide variety 
of long periods as initial guesses.  
We were unable to find any two-planet models that improved $\chi_\nu$ with statistical significance.  
Put another way, with a time baseline of 10\,yr we do not detect curvature 
in the residuals to the one-planet model.

We also considered NLTT\,37349 (2MASS J14264583-0510194)
as the companion responsible for the observed acceleration.  
\cite{Gould04} identified this object as a common proper-motion partner to HD\,126614\,A with a 
sky-projected separation of 42\arcsec.    
The observed photometric properties of this object are $M_V=12.02$, $V = 16.19$, and $V-J=4.00$. 
Presumably NLTT\,37349 has the same parallax (13.8\,mas) and distance (72.4\,pc) as HD\,126614\,A.  
Taken together, these properties are consistent with an M dwarf, roughly M4, 
implying a mass of $m \sim 0.2$\,\msune.  
To compute the approximate gravitational influence of NLTT\,37349, 
we assume that the line of sight separation between HD\,126614\,A and NLTT\,37349 is comparable to 
their physical separation in the sky plane, 
$r_\mathrm{sep} \sim 3000$\,AU.     
The acceleration predicted by this model, 
$\dot{v_r} \sim Gm/r_\mathrm{sep}^2 = 0.004$\,\msye,  
is much too small to account for the observed acceleration of 16.2\,\msye.
Thus, NLTT\,37349 is not the source of the observed linear velocity trend.  

Secular acceleration is a potential cause of a linear velocity trend, 
especially for stars with significant proper motion (such as HD\,126614\,A).  
While our standard velocity pipeline removed this effect from the 
velocities reported Table \ref{tab:rv_126614}, 
it is reassuring to confirm that the calculated secular acceleration 
is inconsistent with the observed velocity trend.
(The pipeline also removed the motion of Keck Observatory about the barycenter of the Solar System.) 
As discussed in \cite{Kurster03} and \cite{Wright09b}, a star's proper motion will cause the 
the radial component of its space velocity vector to change with position on the sky,
resulting a secular acceleration of $\dot{v_r} = D\mu^2$ (to first order),
where $v_r$ is the bulk radial velocity of the star, $D$ is the star's distance, 
and $\mu$ is the total proper motion in radians per unit time.
For the most extreme cases of nearby stars with significant proper motion, 
$\dot{v_r}$ can be as high as a few \msye.
For HD\,126614\,A, we find $\dot{v_r} \sim 0.1$\,\msye, 
ruling out secular acceleration as the cause of the observed 16.2\,\msy trend.

\subsection{AO Observations}
\label{sec:hd126614_ao}

\begin{figure*}
\plotone{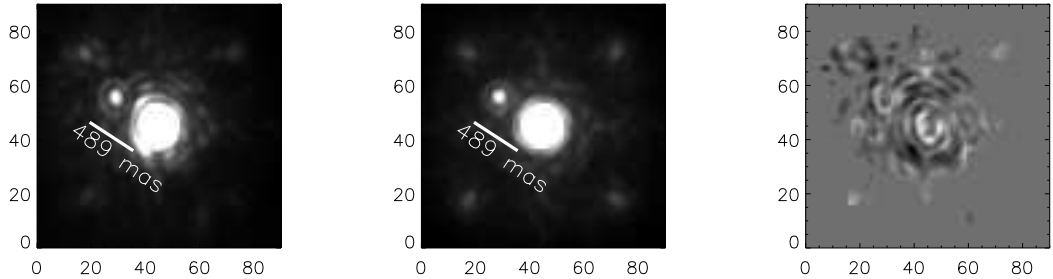}
\caption{Direct observations with adaptive optics of HD\,126614\,A and HD\,126614\,B 
taken with the PHARO imager on Hale 200'' Telescope at Palomar Observatory.
The three panels show K$_s$-band images of 
the target image (left), 
the best two-PSF fit (center), 
and fit residuals (right).  
Each image is oriented North-up and East-left.  
The fainter star, HD\,126614\,B, is clearly detected to the Northeast of the brighter target star, 
HD\,126614\,A.  
The vertical and horizontal axes of each image are labeled with detector pixels 
(the plate scale is 25.2 mas/pixel).  
The image greyscale has been chosen to highlight the companion, Airy rings, and diffraction spikes; 
the residuals (right panel) have been stretched by a factor of 150.  
The full width half-maximum of the AO-corrected PSF was $\sim$100\,mas in each band 
(with Strehl ratios of 30\% in K$_s$-band and 10\% in J-band).
\label{fig:126614_ao}}
\end{figure*}

Having ruled out the above reasons for the linear RV trend, 
we considered a stellar companion close enough to HD\,126614\,A 
to have been missed by prior imaging surveys.  
To search for such a companion, we obtained 
direct imaging of HD\,126614\,A in J, H, and K-short (K$_{\mathrm{s}}$) bands 
on the Hale 200'' Telescope at Palomar Observatory on 13 April
2009 using the facility adaptive optics imager PHARO \citep{Hayward01}.  
In each band, we co-added approximately 150 431-ms exposures.  As expected, the AO
correction was best in K$_{\mathrm{s}}$-band, translating into the smallest errors. 
A nearby faint companion $\sim$500\,mas to the Northeast of HD\,126614\,A 
was identified in each band by visual inspection (Figure \ref{fig:126614_ao}).  
We constructed an empirical PSF in each band from images of a calibrator, 
and images were fit with a two-PSF model.
Parameter errors were calculated using the curvature of
the $\chi^2$ surface, and for consistency, the error in contrast ratio
compared to the percent flux of the fit residuals.  
Combining the astrometric fits in all bands, we find a projected separation of 
489.0 $\pm$ 1.9\,mas at a position angle of 56.1 $\pm$ 0.3\,deg.

We are unaware of a previous detection of this faint companion to HD\,126614\,A, 
which we name HD\,126614\,B.  
We summarize the astrometric and photometric results of the AO observations 
and the inferred properties of HD\,126614\,B in Table~\ref{tab:AO}.

\begin{deluxetable*}{lccccccccc}
\tablecolumns{7} 
\tablecaption{Summary of AO imaging observations and isochrone analysis of HD\,126614\,B.} 
\tablenum{6}
\label{tab:AO}
\tablehead{
\multicolumn{5}{c}{HD\,126614\,B} & & Gl\,661\,B & & Padova Model \\ 
\cline{1-5}  \cline{7-7}  \cline{9-7}\\ 
Filter & Separation & Pos. Angle & Cont. Ratio & $M_{\mathrm{JHK}}$ & & $M_{\mathrm{JHK}}$ & & HD\,126614\,B Mass \\ 
& (mas) & (deg) & & (mag) & & (mag) & & (M$_{\odot}$)  } 
\startdata 
J & 498.7 $\pm$ 6.1 & 57.2 $\pm$ 0.9 & 45.3 $\pm$ 4.1 & 7.42 & & 7.46 & & 0.336 $\pm$ 0.015 \\
H & 487.1 $\pm$ 3.3 & 56.1 $\pm$ 0.6 & 38.8 $\pm$ 1.8 & 6.97 & & 6.97 & & 0.307 $\pm$ 0.007 \\
K$_{\mathrm{s}}$ & 488.5 $\pm$ 2.4 & 56.0 $\pm$ 0.3 & 35.0 $\pm$ 1.1 & 6.75 & & 6.72 & & 0.308 $\pm$ 0.005 \\[5pt]
Combined & 489.0 $\pm$ 1.9 & 56.1 $\pm$ 0.3 & & & & & & 0.324 $\pm$ 0.004\\
\enddata
\tablecomments{
The absolute JHK magnitudes of HD\,126614\,B are
derived from the measured contrast ratios, 2MASS photometry, and
the Hipparcos parallax of HD\,126614\,A.  
The J-band results may be complicated by the presence of a diffraction
spot near the position of the secondary.  The uncertainties listed
are based solely on the uncertainties in the contrast ratios.
Accounting for all uncertainties, we estimate the combined mass to be accurate to perhaps 10\%.
For comparison, we show the absolute JHK magnitudes of Gl\,661\,B, with $M_{\mathrm{V}} = 11.3$.
This star was first identified in \cite{Reid97} by speckle interferometry; 
we use updated photometry from \cite{Reid05}.
}
\end{deluxetable*}

HD\,126614\,A and HD\,126614\,B are unresolved in the spectroscopic and photometric 
observations described in previous sub-sections.
We estimate the amount of V-band contamination from HD\,126614\,B in these observations 
using three methods: model isochrones, observations of a similar star, and
an empirical calibration employing metallicity effects.

First, we interpolated the log(age) = 9.1, $Z$ = 0.070 (corresponding very
nearly to [Fe/H] = +0.56) Padova isochrones \citep{Girardi00} at
these absolute NIR magnitudes.  These interpolations yield mass estimates 
for HD\,126614\,B (Table~\ref{tab:AO}), the average of which is 0.324\,$\pm$\,0.004\,M$_{\odot}$, 
making HD\,126614\,B an M dwarf.  The weighted average V-band
magnitude implied by these measurements is $M_{\mathrm{V}}=10.65 \pm 0.03$.

Second, we find that the M dwarf Gl\,661\,B has nearly identical NIR absolute
magnitudes to HD\,126614\,B (Table~\ref{tab:AO}; \citealt{Reid05}).  
Nonetheless, its absolute V magnitude is
in severe disagreement with the $Z$ = 0.070 isochrones (and in even worse
agreement with the solar metallicity isochrones).  This suggests that
the Padova isochrones significantly overestimate the V-band flux of M
dwarfs.  We thus consider the isochrone V magnitudes to be unreliable.

Finally, we have applied the M dwarf photometric calibration of
\cite{Johnjohn09a} to the absolute K$_{\mathrm{s}}$ magnitude derived from the
AO photometry.  This empirical calibration is based on G--M binaries with
well-measured metallicities (from LTE spectral synthesis analysis of
the primary) and K$_{\mathrm{s}}$-band magnitudes of both components.  Johnson \&
Apps find that a star with [Fe/H] = +0.56 and $M_{\mathrm{K_{s}}}$ = 6.75
should have $M_{\mathrm{V}}=12.2$.  
This estimate, which we feel is the most robust of the three we have
performed, implies V = 16.5 and therefore negligible contamination in
our optical spectra ($\Delta$V = 7.8, $<$0.1\% contamination).

Knowledge of the mass of HD\,126614\,B from the isochrones combined with
the observed RV trend of HD\,126614\,A allows us to constrain the true physical
separation between the stars.  Let $\theta$ be the angle between
the line of sight to the primary and the line connecting the primary
to the secondary, such that the secondary is behind the primary when $\theta=0$.  
The observed radial velocity trend (the instantaneous acceleration along the line of sight) 
is related to $\theta$, the mass of the
secondary $M_\mathrm{B}$, and the true physical separation $r_\mathrm{AB}$ by 
\begin{equation}
\label{eq:v_dot}
\dot{v_\mathrm{A}}=\frac{GM_\mathrm{B}}{r_\mathrm{AB}^2}\cos\theta,
\end{equation}
where $r_\mathrm{AB}$ is related to the apparent angular separation $\rho_\mathrm{AB}$ through
\begin{equation}
\label{eq:v_dot2}
\left(\frac{r_\mathrm{AB}}{\mathrm{AU}}\right)\sin\theta = \frac{\rho_\mathrm{AB}}{\pi_\mathrm{A}} 
\end{equation}
and $\pi_\mathrm{A}$, the Hipparcos parallax of HD\,126614\,A.  
Given the observed radial acceleration of 16.2\,\msy  
and the best separation and mass from
Table~\ref{tab:AO}, there are two solutions for the implied physical
separation between the stars:  
$r_\mathrm{AB}$ = $40^{+7}_{-4}$ and $50^{+2}_{-3}$\,AU
(where the uncertainty is dominated by the uncertainty in the parallax),
which we can crudely combine as $45^{+7}_{-9}$\,AU with the caveat that
the probability distribution function is highly non-Gaussian.

\cite{Holman99} studied the dynamical stability of planetary
orbits in binary systems.  In the terms of that work, we find the mass
ratio of the binary system to be $\mu = 0.22$, at the edge of Hill's
Regime.  Holman \& Wiegert find that for a binary in a circular orbit,
planets in S-type orbits of semimajor axis $a_b$ are generally stable
when secondary stars orbit with semimajor axes $a_\mathrm{B} > a_b / 0.38$.
For our system, this yields $a_\mathrm{B} > 6.2$\,AU, consistent with our AO
astrometry.  Indeed, the HD\,126614 system unconditionally passes
this weak test for stability for $e_\mathrm{B} < 0.6$; beyond this regime more
detailed knowledge of the binary orbit is required for further
analysis.

\cite{Pfahl06} discuss the prevalence of planets in
binary systems such as HD\,126614. They note that the literature
contains a few planet detections in systems separated by $\sim 20$\,AU,
the most similar to HD\,126614 being HD\,196885 (Fischer et al. 2009,
submitted).
Thus, HD\,126614\,Ab appears to be in a stable orbit 
in a binary system that is not atypical, even among systems with detected planets.  

\subsection{Planetary Interpretation}
\label{sec:126614_pi}

The identification of HD\,126614 as a stellar binary with a 
separation of $\sim$0\farcs5 (unresolved by Keck/HIRES) 
raises the question of the origin of the periodic RV signal described in 
\S\,\ref{sec:126614_RV}.  
We investigated the possibility that the RV variation was the result of 
distortions in the spectral line profiles due to contamination of 
the HD\,126614\,A spectrum by the much fainter HD\,126614\,B spectrum 
modulated by an orbital companion of its own.
We present multiple lines of reasoning to argue against this alternative explanation 
and in favor of the Jovian-mass planet orbiging HD\,126614\,A 
described in \S\,\ref{sec:126614_RV}.

From the AO observations in JHK bands, we estimated the V-band flux 
of HD\,126614\,B in several ways with the most reliable method 
giving a flux ratio (A/B) of 7.8\,mag.  
Thus in the iodine region of the spectrum, the lines from HD\,126614\,B 
are fainter by a factor of $\sim$$10^3$.   
To produce a Doppler signal with $K$\,=\,7\,\ms when diluted by HD\,126614\,A,  
the the signal from HD\,126614\,B would have an amplitude of approximately $K$\,=\,7\,\ks 
(implying a companion orbiting HD\,126614\,B with mass 0.25\,\msune, an M dwarf).  
Yet, the stellar lines of HD\,126614\,A, with \vsini\,=\,1.52\,\kse, are narrower 
than the amplitude of the hypothetical $K$\,=\,7\,\ks Doppler signal.  
Thus, the stellar line profiles of HD\,126614\,A and HD\,126614\,B would not be blended, 
but separated from each other for most of the 3.41\,yr orbit.  
This inconsistency casts serious doubt on the blend scenario 
as an explanation for the observed RV variation.  

Nevertheless, we investigated the blend scenario further with a spectral line bisector span test
\citep{Torres05,Queloz2001}.  
We chose a region of the spectrum (6795--6865\,\AA) to the red of the iodine region for 
increased sensitivity to the spectral features of HD\,126614\,B (an M dwarf) 
and a lack of telluric contamination.  
We cross-correlated each post-upgrade Keck spectrum against the solar spectrum using the 
National Solar Observatory solar atlas \citep{NSO_atlas}.  
From this representation of the average spectral line profile we computed the bisectors 
of the cross-corrleation peak, 
and as a measure of the line asymmetry we calculated the ``bisector span'' as the velocity 
difference between points selected near the top and bottom cross-correlation peak.  
If the velocities were the result of a blend, we would expect the 
line bisectors to vary in phase with the 3.41\,yr period with an amplitude similar to K\,=\,7\,\mse.  
Instead, we find no significant correlation between the bisector spans and the RVs 
after subtracting the 16.2\,\msy trend.  
The Pearson linear correlation coefficient between these two quantities, $r$\,=\,$-$0.28, 
demonstrates the lack of correlation.  
We conclude that the velocity variations are real and that the star is orbited by a 
Jovian planet. 



\section{HD\,13931}
\label{sec:hd13931}

\subsection{Stellar Characteristics}

HD\,13931 (=\,HIP\,10626) is spectral type G0, with $V$\,=\,7.61, $B$\,--\,$V$\,=\,0.64, and 
$M_V$\,=\,4.32, placing it $\sim$0.4 mag above the main sequence.   
It is chromospherically quiet with $S_\mathrm{HK}$\,=\,0.16 and \lrphk\,=\,$-$4.99 \citep{Isaacson09}, 
implying a rotation period of $\sim$26 days \citep{Noyes84}.
\cite{Valenti05} measured an approximately solar metallicity of \feh\,=\,+0.03 using SME.  
Its placement above the main sequence and low chromospheric activity are 
consistent with an old, quiescent star of age 6.4--10.4\,Gyr.

\subsection{Photometry from Fairborn Observatory}

We have 181 photometric observations of HD\,13931 over two consecutive 
observing seasons.  Periodogram analyses found no significant periodicity 
over the range 1--100 days.  The short-term standard deviations of the 
target and comparison stars (Table~\ref{tab:photometry}) provide an upper limit of 0.0016 mag 
for night-to-night variability.  The long-term standard deviations suggest 
intrinsic low-level variability in HD\,13931, but additional observations 
are needed to confirm it.

\subsection{Doppler Observations and Keplerian Fit}

\begin{figure}
\epsscale{1.2}
\plotone{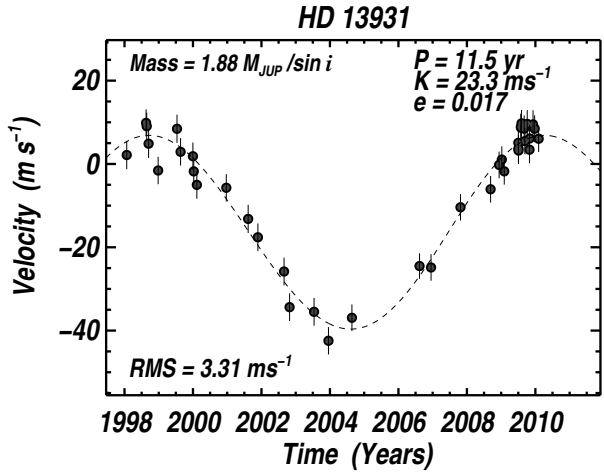}
\caption{Measured velocity vs.\ time for HD\,13191 (\textit{filled circles})
    with the associated best-fit Keplerian model (\textit{dashed line}).  
    The error bars show the quadrature sum of measurement uncertainties and jitter.}
\label{fig:13931_ts_1p_1}
\end{figure}

We monitored the radial velocity of HD\,13931 at Keck for the past 12\,yr.
Table \ref{tab:rv_13931} gives the Julian Dates of the observations, 
the measured (relative) radial velocities, and the associated measurement uncertainties (excluding jitter).  
The 39 velocities listed in Table \ref{tab:rv_13931} have an rms of 15.1\,\mse.
These velocities are plotted as a time series in Figure \ref{fig:13931_ts_1p_1} 
along with the best-fit Keplerian model.

A long period signal that has completed approximately one cycle is clearly seen in 
the velocities plotted in Figure \ref{fig:13931_ts_1p_1}.  
After trying a wide variety of trial periods, we found the best-fit, 
single-planet Keplerian model with   
$P$\,=\,11.5\,$\pm$\,1.1\,yr, 
$e$\,=\,0.02\,$\pm$\,0.08, and 
$K$\,=\,23.3\,$\pm$\,2.9\,\mse,
implying a planet of minimum mass $M$\,sin\,$i$\,=\,1.88\,\mjup 
orbiting with semimajor axis $a$\,=\,5.15\,AU.   
Note that the orbit is consistent with circular.  
The full set of orbital parameters is listed in Table \ref{tab:orbital_params}.
This model, with $\chi_{\nu}$\,=\,1.01 and 3.31\,\ms rms velocity residuals, 
is an excellent fit to the data.
Under the assumption of a one-planet model, 
we observed three RV extrema with Keck that strongly constrain the orbital parameters.

\section{Gl\,179}
\label{sec:hip22627}

\subsection{Stellar Characteristics}

Gl\,179 (=\,HIP\,22627) is spectral type M3.5, with $V$\,=\,11.96, $B$\,--\,$V$\,=\,1.59, and 
$M_V$\,=\,11.5, placing it $\sim$0.3 mag above the main sequence.   
It is chromospherically active with $S_\mathrm{HK}$\,=\,0.96 \citep{Isaacson09}.  
\cite{Valenti05} did not calculate stellar parameters using SME for Gl\,179 because of its late spectral type.
Similarly, the \cite{Noyes84} calibration of stellar rotation does not apply for $B-V$\,$>$\,1.4.  
\cite{Johnjohn09a} find a high metallicity of \feh\,=\,0.30\,$\pm$\,0.10 
for Gl\,179 based on its absolute K-band magnitude, $M_K$, and $V-K$ color.

\subsection{Photometry from Fairborn Observatory}

We have not made photometric observations of Gl\,179 from the APTs because at 
$V$\,=\,11.96 the star is too faint.
 
\subsection{Doppler Observations and Keplerian Fit}

\begin{figure}
\epsscale{1.2}
\plotone{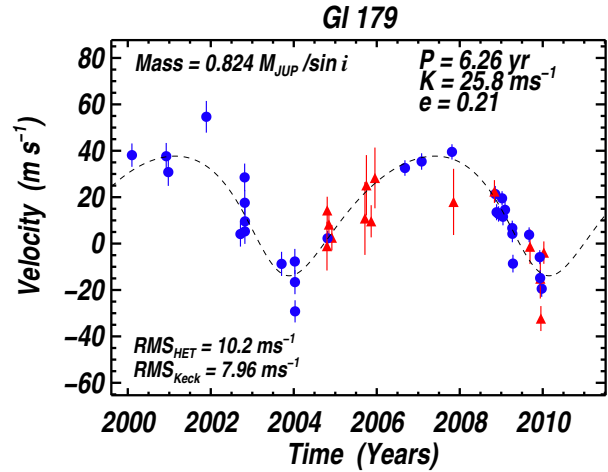}
\caption{Measured velocity vs.\ time for Gl\,179 (\textit{filled circles})
    with the associated best-fit Keplerian model (\textit{dashed line}).
    Filled blue circles represent measurements from Keck, 
    while filled red triangles represent HET measurements.
    The error bars show the quadrature sum of measurement uncertainties and jitter.}
\label{fig:hip22627_ts_1p_1}
\end{figure}

We monitored the radial velocity of Gl\,179 at Keck for the past 10\,yr 
and at the HET for the past 5\,yr.
Table \ref{tab:rv_hip22627} gives the Julian Dates of the observations, 
the measured (relative) radial velocities, the associated measurement uncertainties (excluding jitter), 
and the telescope used.  
The 44 velocities listed in Table \ref{tab:rv_hip22627} have an rms of 19.7\,\mse.
Because Gl\,179 is a faint M dwarf, the measurements have lower signal-to-noise ratios 
and larger uncertainties compared to the other stars in this paper.  
The measured velocities are plotted as a time series in Figure \ref{fig:hip22627_ts_1p_1} 
along with the best-fit Keplerian model.

After trying a wide variety of trial orbital periods, we find the best-fit, 
single-planet Keplerian model has  
$P$\,=\,6.26\,$\pm$\,0.16\,yr, 
$e$\,=\,0.21\,$\pm$\,0.08, and 
$K$\,=\,25.8\,$\pm$\,2.2\,\mse,
implying a planet of minimum mass $M$\,sin\,$i$\,=\,0.82\,\mjup 
orbiting with semimajor axis $a$\,=\,2.41\,AU.
The full set of orbital parameters is listed in Table \ref{tab:orbital_params}.  
We allowed for floating offsets between the HET RVs, the pre-upgrade Keck RVs, 
and the post-upgrade RVs when fitting the data.

Note that the planet is clearly detected in the Keck RVs alone, 
but the HET measurements are crucial for the precise determination of orbital parameters, 
especially eccentricity and minimum mass.  
This single planet model, with $\chi_{\nu}$\,=\,1.78 and 9.51\,\ms rms velocity residuals, 
is a statistically significant improvement over a model with no planets.  
However, the relatively large value of $\chi_{\nu}$ implies that either 
additional planets remain undetected in the system
or we have underestimated the measurement uncertainties or jitter.  
We will continue to observe Gl\,179 to search for additional planets.

\section{Discussion}
\label{sec:discussion}

We present the detection of four new extrasolar planets.  
These planets add to the statistical properties of giant planets in long-period orbits near the ice line.
Long observational time baselines (10--12\,yr) were necessary to accurately measure the 
high-eccentricity, low-amplitude signals of HD\,34445b and HD\,126614\,Ab, 
as well as the intermediate-amplitude, but long-period signals of HD\,13931\,b and Gl\,179\,b.

HD\,34445\,b is a massive planet ($M$\,sin\,$i$\,=\,0.79\,\mjupe) in a mildly eccentric ($e$\,=\,0.27), 
long-period ($P$\,=\,2.87\,yr) orbit around an old G0 star.  
We clearly detect this planet in the Keck and HET data sets individually, 
and their combination allows for more precise orbit determination.
The relatively large residuals to the one-planet fit (rms = 6--8\,\mse) 
hint at a second, unresolved planet in the system.  
Underestimated jitter would also explain the large residuals, 
but we deem this explanation less likely given 
the metallicity, color, and modest evolution of HD\,34445.

HD\,126614\,Ab is a massive planet ($M$\,sin\,$i$\,=\,0.38\,\mjupe) 
in a long-period ($P$\,=\,3.41\,yr), eccentric ($e$\,=\,0.41) orbit around an extremely metal-rich star.  
At \feh\,=\,+0.56\,$\pm$0.04, HD\,126614\,A has the highest metallicity of the 1040 stars in the 
SPOCS catalog \citep{Valenti05}.
It also has the highest \feh\ of the $\sim$250 stars with known planets and measured metallicities. 
We confirmed the high metallicity of HD\,126614\,A by running a separate iodine-free HIRES spectrum 
through the same SME pipeline used for the SPOCS catalog.
We found \feh\,=\,+0.51\,$\pm$0.04, consistent with the SPOCS catalog value.
Other authors also find an extremely high metallicity, 
including \cite{Castro1997} and \cite{Cenarro07}, who both find \feh\,=\,+0.55. 
These measurements, along with the detection of HD\,126614\,Ab, 
add statistical weight to the strong positive correlation between 
giant planet occurrence and metallicity \citep{Fischer05}.  
Indeed, HD\,126614\,A has been part of the planet-metallicity correlation story for some time.
In an early paper discussing the host star properties of some of the first extrasolar planets, 
\cite{Gonzalez99} suggested that two bright, high-metallicity stars, 
namely HD\,99109 and HD\,126614, should be searched for Doppler variations.  
HD\,99109 is known to host a planet with minimum mass $M$\,sin\,$i$\,=\,0.5\,\mjup and 
orbital period $P$\,=\,1.2\,yr \citep{Butler06}, 
and HD\,126614\,A now joins the list of high-metallity stars with planets.

In addition to the planet orbiting HD\,126614\,A, we detected a faint M dwarf companion 
using adaptive optics and the PHARO camera at Palomar Observatory.    
This previously undiscovered star, HD\,126614\,B, has an estimated mass of 0.32\,M$_{\odot}$ and 
is separated from HD\,126614\,A by 489\,$\pm$\,1.9\,mas at position angle 56.1\,$\pm$\,0.3\,deg.
This corresponds to a projected separation of 33\,AU.

HD\,13931\,b is reminiscent of Jupiter in orbital period ($P$\,=\,11.5\,yr), 
eccentricity ($e$\,=\,0.02), and to a lesser extent mass ($M$\,sin\,$i$\,=\,1.88\,\mjupe).  
The host star, HD\,13931, is also similar to the Sun in mass ($M_{\star}$\,=\,1.02\,$M_{\sun}$) 
and metallicity (\feh\,=\,+0.03).  
HD\,13931\,b is one of only 4 known RV-detected planets with orbital periods longer than 10\,yr.  
The other such planets---55\,Cnc\,d \citep{Fischer08},
HD\,217107\,c \citep{Vogt05,Wright09}
and HD\,187123\,c \citep{Wright07,Wright09}---are all in multi-planet systems.

Gl\,179\,b is a Jovian-mass ($M$\,sin\,$i$\,=\,0.82\,\mjupe) planet in a long-period ($P$\,=\,6.3\,yr) orbit.  
The host star, Gl\,179, is one of only $\sim$10 M dwarfs currently known to host a planet 
and is among the faintest ($V$\,=\,11.96) stars with a planet discovered by RV measurements.  
This planet is detected in the Keck velocities alone, but without the HET measurements, 
the orbital parameters, especially eccentricity and minimum mass, 
would be determined more poorly.

We note with interest that Gl\,179 is an almost identical twin to Gl\,876, 
an M dwarf known to host two Jovian planets locked in resonant orbits  
and a super-Earth in a $P$\,=\,1.9\,d orbit \citep{Marcy_876,Rivera05}.  
The stars are similar in effective temperature, mass, and age, 
as traced respectively by $V-K$ color (5.00 for Gl\,179 and 5.15 for Gl\,876), 
$M_K$ (6.49 for Gl\,179 and 6.67 for Gl\,876), and 
$S_\mathrm{HK}$ (0.96 for Gl\,179 and 1.02 for Gl\,876).  
The high metallicity of Gl\,179 is also strikingly similar to the metallicity of Gl\,876.  
\cite{Johnjohn09a} estimate \feh\ = $+$0.3 and $+$0.37 for Gl\,179 and Gl\,876, respectively. 
Based on their analysis of M dwarfs with planets, 
Johnson \& Apps find that the planet-metallicity correlation holds for stars 
at the bottom of the main-sequence. 
Gl\,179 and its planet add statistical weight to this finding.

The planet-bearing stars presented here are good candidates for follow-up 
astrometric and direct imaging observations.  
The astrometric perturbations from these planets on their host stars 
(10's to 100's of $\mu$as) will be quite detectable with the expected sub-$\mu$as 
sensitivity of NASA's Space Interferometry Mission (SIM) or its proposed variants.  
Direct detection is also plausible using coronagraphs/interferometers 
on space-borne and next-generation ground-based observatories, including GPI and SPHERE.
HD\,13931\,b may be the best candidate with a maximum projected angular separation of 120\,mas.
Indeed, SIM observations of HD\,13931, 
even over a fraction of the orbital period (SIM has a planned mission duration of 5 years),
combined with the RV measurements presented here,  
would completely determine the 3-dimensional orbit, 
giving accurate predictions of the best times for direct imaging observations.

\acknowledgments{We thank the many observers who contributed to the velocities reported here.  
We gratefully acknowledge the efforts and dedication of the Keck Observatory staff, 
especially Grant Hill and Scott Dahm for support of HIRES 
and Greg Wirth for support of remote observing.  
We are grateful to the time assignment committees of the University of California, NASA, and NOAO  
for their generous allocations of observing time.  
Without their long-term commitment to radial velocity monitoring, 
these long-period planets would likely remain unknown.  
We acknowledge R.\ Paul Butler and S.\ S.\ Vogt for many years
of contributing to the data presented here.
A.\,W.\,H.\ gratefully acknowledges support from a Townes Post-doctoral Fellowship 
at the U.\,C.\ Berkeley Space Sciences Laboratory.
J.\,A.\,J.\ is an NSF Astronomy and Astrophysics Postdoctoral Fellow and
acknowledges support form NSF grant AST-0702821.
G.\,W.\,M.\ acknowledges NASA grant NNX06AH52G.  
J.\,T.\,W.\ received support from NSF grant AST-0504874.
M.\,E. and W.\,D.\,C acknowledge support from NASA  
grants NNX07AL70G and NNX09AB30G issued through the Origins of Solar Systems Program.
Automated Astronomy at Tennessee State University has been supported by
NASA and NSF as well as Tennessee State University and the State of
Tennessee through its Centers of Excellence program.
This work made use of the SIMBAD database (operated at CDS, Strasbourg, France), 
NASA's Astrophysics Data System Bibliographic Services, 
and the NASA Star and Exoplanet Database (NStED).
Finally, the authors wish to extend special thanks to those of Hawai`ian ancestry 
on whose sacred mountain of Mauna Kea we are privileged to be guests.  
Without their generous hospitality, the Keck observations presented herein
would not have been possible.}

\bibliographystyle{apj}
\bibliography{sixpack}


\clearpage
\LongTables 

\begin{deluxetable}{cccc}
\tabletypesize{\footnotesize}
\tablecaption{Radial Velocities of HD\,34445}
\tablenum{4}
\label{tab:rv_34445}
\tablewidth{0pt}
\tablehead{ 
    \colhead{JD -- 2440000}            & \colhead{Radial Velocity}  & \colhead{Uncertainty} & \colhead{Telescope }  \\
    \colhead{}  & \colhead{(\mse)}   & \colhead{(\mse)}  & \colhead{}}
\startdata
10838.76212 &  -20.59 &    1.83  &              K                     \\ 
 11051.10569 &   -1.84 &    1.40  &              K                     \\ 
 11073.04246 &   -2.97 &    1.21  &              K                     \\ 
 11171.84795 &    3.84 &    1.55  &              K                     \\ 
 11228.80594 &    2.02 &    1.47  &              K                     \\ 
 11550.88478 &    5.75 &    1.35  &              K                     \\ 
 11551.88096 &    3.79 &    1.43  &              K                     \\ 
 11581.87017 &    9.09 &    1.54  &              K                     \\ 
 11898.03487 &  -20.63 &    1.39  &              K                     \\ 
 11974.76461 &   -2.35 &    1.53  &              K                     \\ 
 12003.74221 &  -13.51 &    1.50  &              K                     \\ 
 12007.72006 &   -8.73 &    1.41  &              K                     \\ 
 12188.14432 &    5.87 &    1.59  &              K                     \\ 
 12219.15109 &   13.75 &    1.79  &              K                     \\ 
 12220.08113 &    0.52 &    1.80  &              K                     \\ 
 12235.86269 &   -3.92 &    1.38  &              K                     \\ 
 12238.88934 &    3.14 &    1.61  &              K                     \\ 
 12242.92553 &    6.49 &    1.39  &              K                     \\ 
 12572.99576 &   21.87 &    1.50  &              K                     \\ 
 12651.93917 &    2.68 &    1.53  &              K                     \\ 
 12899.09825 &  -16.22 &    1.29  &              K                     \\ 
 12926.86832 &  -23.82 &    4.99  &              H                     \\ 
 12932.00108 &  -22.46 &    2.09  &              H                     \\ 
 12940.99257 &  -18.74 &    3.23  &              H                     \\ 
 12942.98461 &  -26.21 &    2.37  &              H                     \\ 
 12978.73611 &  -24.64 &    4.03  &              H                     \\ 
 12979.73953 &  -22.54 &    4.01  &              H                     \\ 
 12983.88276 &  -13.97 &    2.21  &              H                     \\ 
 12984.88936 &  -20.95 &    3.41  &              H                     \\ 
 12986.73101 &   -6.99 &    4.35  &              H                     \\ 
 12986.86187 &  -14.40 &    2.99  &              H                     \\ 
 13016.88898 &    1.63 &    1.21  &              K                     \\ 
 13044.79245 &   -6.83 &    1.78  &              K                     \\ 
 13255.96162 &   14.40 &    3.08  &              H                     \\ 
 13258.95666 &   12.95 &    3.01  &              H                     \\ 
 13260.96242 &   15.60 &    2.06  &              H                     \\ 
 13262.94700 &   12.35 &    1.76  &              H                     \\ 
 13286.88668 &    8.19 &    3.05  &              H                     \\ 
 13303.12110 &   11.45 &    1.35  &              K                     \\ 
 13338.90162 &   -1.81 &    1.21  &              K                     \\ 
 13340.02404 &   -3.95 &    1.26  &              K                     \\ 
 13359.69356 &    8.66 &    3.28  &              H                     \\ 
 13365.68179 &   12.38 &    2.36  &              H                     \\ 
 13368.95279 &    5.50 &    1.16  &              K                     \\ 
 13369.80021 &    2.74 &    1.19  &              K                     \\ 
 13756.75932 &    7.60 &    3.08  &              H                     \\ 
 13775.69673 &   -2.11 &    4.85  &              H                     \\ 
 13780.69018 &    6.51 &    2.08  &              H                     \\ 
 13841.76181 &   -0.41 &    1.05  &              K                     \\ 
 13978.97450 &  -13.27 &    2.42  &              H                     \\ 
 13979.97609 &  -19.30 &    1.56  &              H                     \\ 
 13982.10169 &  -14.37 &    1.16  &              K                     \\ 
 13983.09819 &   -8.77 &    0.73  &              K                     \\ 
 13984.10344 &   -8.36 &    0.89  &              K                     \\ 
 13985.04721 &  -10.67 &    1.11  &              K                     \\ 
 13985.98820 &   -7.28 &    3.29  &              H                     \\ 
 13988.96142 &  -12.49 &    3.31  &              H                     \\ 
 14003.92771 &  -14.39 &    1.91  &              H                     \\ 
 14043.81148 &  -23.50 &    3.20  &              H                     \\ 
 14055.77338 &  -15.55 &    2.73  &              H                     \\ 
 14071.89776 &   -6.16 &    3.70  &              H                     \\ 
 14096.67373 &   -1.56 &    3.40  &              H                     \\ 
 14121.60247 &   -0.97 &    3.34  &              H                     \\ 
 14130.87196 &   -0.18 &    1.17  &              K                     \\ 
 14181.59907 &    0.66 &    5.45  &              H                     \\ 
 14344.11108 &   14.59 &    1.16  &              K                     \\ 
 14346.97829 &    4.11 &    1.86  &              H                     \\ 
 14347.96596 &   11.32 &    3.26  &              H                     \\ 
 14377.89952 &   12.25 &    1.64  &              H                     \\ 
 14398.06266 &    4.62 &    1.17  &              K                     \\ 
 14419.95414 &   -5.91 &    3.91  &              H                     \\ 
 14452.85410 &   -1.18 &    2.39  &              H                     \\ 
 14475.63706 &    1.73 &    3.76  &              H                     \\ 
 14492.85216 &    6.39 &    1.25  &              K                     \\ 
 14500.70733 &    3.84 &    2.86  &              H                     \\ 
 14544.81179 &   17.87 &    1.41  &              K                     \\ 
 14546.73807 &   19.12 &    1.16  &              K                     \\ 
 14547.60716 &   19.93 &    3.26  &              H                     \\ 
 14730.92837 &    4.20 &    4.83  &              H                     \\ 
 14762.00435 &    3.81 &    3.93  &              H                     \\ 
 14777.95271 &    4.68 &    1.02  &              K                     \\ 
 14781.79952 &    7.11 &    1.62  &              H                     \\ 
 14791.00635 &   15.34 &    1.26  &              K                     \\ 
 14807.92630 &   -4.41 &    1.46  &              K                     \\ 
 14810.89575 &   -0.73 &    1.37  &              K                     \\ 
 14838.84447 &  -15.41 &    1.40  &              K                     \\ 
 14846.97512 &  -11.01 &    1.61  &              K                     \\ 
 14856.59752 &  -15.94 &    2.22  &              H                     \\ 
 14864.89937 &  -19.53 &    1.24  &              K                     \\ 
 14865.83319 &  -18.59 &    0.96  &              K                     \\ 
 14867.78335 &  -21.11 &    1.35  &              K                     \\ 
 14877.68465 &  -16.60 &    2.26  &              H                     \\ 
 14927.76208 &  -12.33 &    1.36  &              K                     \\ 
 14928.72560 &  -15.91 &    1.23  &              K                     \\ 
 14929.73178 &   -7.89 &    0.92  &              K                     \\ 
 14934.73601 &   -7.39 &    1.36  &              K                     \\ 
 15045.13099 &  -11.01 &    1.59  &              K                     \\ 
 15049.12654 &  -21.37 &    1.43  &              K                     \\ 
 15074.98600 &  -24.11 &    4.74  &              H                     \\ 
 15076.12099 &  -23.84 &    2.69  &              K                     \\ 
 15080.13683 &  -25.10 &    2.79  &              K                     \\ 
 15084.13106 &  -13.69 &    2.79  &              K                     \\ 
 15101.90535 &  -34.62 &    5.14  &              H                     \\ 
 15110.12260 &  -21.65 &    1.46  &              K                     \\ 
 15123.00538 &   -6.66 &    5.95  &              H                     \\ 
 15135.10186 &  -11.30 &    1.37  &              K                     \\ 
 15135.95876 &  -10.74 &    1.39  &              K                     \\ 
 15142.80591 &  -16.77 &    4.28  &              H                     \\ 
 15170.83820 &   -8.05 &    1.29  &              K                     \\ 
 15187.83253 &   -3.26 &    1.27  &              K                     \\ 
 15189.83159 &   -6.05 &    1.37  &              K                     \\ 
 15190.67571 &   -5.15 &    4.38  &              H                     \\ 
 15197.82141 &  -11.12 &    1.38  &              K                     \\ 
 15215.60340 &  -11.95 &    2.12  &              H                     \\ 
 15229.77934 &   -5.58 &    1.21  &              K                     \\ 
 15231.95374 &   -7.91 &    1.21  &              K                     \\ 
 15251.85907 &   -6.17 &    1.43  &              K                     \\ 
 15255.79845 &   -3.31 &    1.39  &              K                     \\ 
\enddata
\end{deluxetable}

\begin{deluxetable}{cccc}
\tabletypesize{\footnotesize}
\tablecaption{Radial Velocities of HD\,126614\,A}
\tablenum{5}
\label{tab:rv_126614}
\tablewidth{0pt}
\tablehead{ 
    \colhead{JD -- 2440000}               & \colhead{Radial Velocity}  & \colhead{Uncertainty} & \colhead{Telescope}  \\
    \colhead{}  & \colhead{(\ms)}            & \colhead{(\ms)}       & \colhead{}
}
\startdata
 11200.13355 & -175.78 &    1.26  &              K                     \\ 
 11311.92294 & -169.56 &    2.17  &              K                     \\ 
 11342.85107 & -155.44 &    1.47  &              K                     \\ 
 11370.81727 & -151.70 &    1.44  &              K                     \\ 
 11373.83647 & -146.70 &    1.77  &              K                     \\ 
 11552.15999 & -135.05 &    1.28  &              K                     \\ 
 11553.16885 & -136.56 &    1.34  &              K                     \\ 
 11581.17426 & -134.75 &    1.26  &              K                     \\ 
 11583.06953 & -140.19 &    1.44  &              K                     \\ 
 11585.12509 & -138.05 &    1.07  &              K                     \\ 
 11586.03518 & -134.02 &    1.14  &              K                     \\ 
 11680.01873 & -141.32 &    1.54  &              K                     \\ 
 11982.14743 & -126.57 &    1.45  &              K                     \\ 
 12003.94759 & -117.29 &    1.48  &              K                     \\ 
 12005.13470 & -118.83 &    1.38  &              K                     \\ 
 12065.93845 & -123.18 &    1.49  &              K                     \\ 
 12096.76464 & -116.69 &    1.78  &              K                     \\ 
 12128.75811 & -119.29 &    1.23  &              K                     \\ 
 12335.07744 & -112.03 &    1.39  &              K                     \\ 
 12363.08412 & -108.95 &    1.52  &              K                     \\ 
 12389.99741 & -110.09 &    1.77  &              K                     \\ 
 12445.84007 & -107.11 &    1.37  &              K                     \\ 
 12683.05902 &  -82.87 &    1.34  &              K                     \\ 
 12711.99652 &  -81.84 &    1.40  &              K                     \\ 
 12776.95616 &  -76.98 &    1.47  &              K                     \\ 
 12849.80405 &  -78.59 &    1.51  &              K                     \\ 
 13015.15707 &  -70.77 &    1.29  &              K                     \\ 
 13016.16478 &  -69.55 &    1.20  &              K                     \\ 
 13017.15230 &  -70.34 &    1.25  &              K                     \\ 
 13179.83922 &  -66.55 &    1.47  &              K                     \\ 
 13425.13398 &  -57.71 &    1.18  &              K                     \\ 
 13777.11170 &  -41.41 &    1.02  &              K                     \\ 
 13838.01968 &  -39.62 &    1.62  &              K                     \\ 
 14131.15530 &  -15.75 &    1.24  &              K                     \\ 
 14246.94875 &  -18.96 &    0.93  &              K                     \\ 
 14247.94129 &  -18.73 &    1.21  &              K                     \\ 
 14248.90277 &  -18.55 &    1.16  &              K                     \\ 
 14251.83500 &  -23.10 &    1.13  &              K                     \\ 
 14255.78231 &  -20.33 &    0.87  &              K                     \\ 
 14277.75659 &  -19.17 &    1.13  &              K                     \\ 
 14278.76238 &  -18.63 &    1.22  &              K                     \\ 
 14285.77622 &   -9.74 &    1.28  &              K                     \\ 
 14294.83121 &  -18.26 &    1.19  &              K                     \\ 
 14300.80378 &  -16.35 &    1.15  &              K                     \\ 
 14493.17109 &   -9.46 &    1.12  &              K                     \\ 
 14549.01403 &  -10.28 &    1.26  &              K                     \\ 
 14639.85723 &  -11.05 &    1.25  &              K                     \\ 
 14839.12869 &   -1.44 &    0.87  &              K                     \\ 
 14865.17003 &   -3.63 &    1.66  &              K                     \\ 
 14866.08003 &   -2.67 &    0.91  &              K                     \\ 
 14927.91558 &   -3.71 &    1.42  &              K                     \\ 
 14929.09501 &   -3.22 &    1.31  &              K                     \\ 
 14930.03891 &   -2.54 &    1.15  &              K                     \\ 
 14964.02038 &   -2.01 &    0.73  &              K                     \\ 
 14983.79390 &    1.65 &    1.23  &              K                     \\ 
 14984.84861 &    4.14 &    1.19  &              K                     \\ 
 14985.95010 &    2.85 &    1.38  &              K                     \\ 
 14986.93908 &    7.48 &    1.23  &              K                     \\ 
 14987.94064 &    4.43 &    1.30  &              K                     \\ 
 14988.95513 &    1.50 &    1.28  &              K                     \\ 
 15014.82834 &    3.02 &    1.24  &              K                     \\ 
 15015.83799 &   -1.02 &    1.25  &              K                     \\ 
 15016.88768 &    0.91 &    1.07  &              K                     \\ 
 15041.84683 &    4.80 &    1.33  &              K                     \\ 
 15042.87063 &    8.59 &    1.45  &              K                     \\ 
 15043.81939 &    3.67 &    1.33  &              K                     \\ 
 15044.79615 &    7.78 &    1.40  &              K                     \\ 
 15190.15992 &   19.99 &    1.34  &              K                     \\ 
 15197.15781 &   21.27 &    1.23  &              K                     \\ 
 15229.04986 &   27.36 &    1.19  &              K                     \\ 
\enddata
\end{deluxetable}

\begin{deluxetable}{cccc}
\tabletypesize{\footnotesize}
\tablecaption{Radial Velocities of HD\,13931}
\tablenum{7}
\label{tab:rv_13931}
\tablewidth{0pt}
\tablehead{ 
    \colhead{JD -- 2440000}               & \colhead{Radial Velocity}  & \colhead{Uncertainty} & \colhead{Telescope}  \\
    \colhead{}  & \colhead{(\ms)}            & \colhead{(\ms)}       & \colhead{}
}
\startdata
 10837.81679 &    2.15 &    1.59  &              K                     \\ 
 11044.11226 &    9.86 &    1.26  &              K                     \\ 
 11051.05061 &    9.06 &    1.27  &              K                     \\ 
 11071.05334 &    4.87 &    1.66  &              K                     \\ 
 11172.82617 &   -1.57 &    1.34  &              K                     \\ 
 11374.13082 &    8.45 &    1.55  &              K                     \\ 
 11412.08069 &    2.92 &    1.37  &              K                     \\ 
 11543.80646 &    1.85 &    1.38  &              K                     \\ 
 11552.79058 &   -1.78 &    1.56  &              K                     \\ 
 11585.71983 &   -5.01 &    1.46  &              K                     \\ 
 11899.88418 &   -5.72 &    1.33  &              K                     \\ 
 12133.12439 &  -13.19 &    1.56  &              K                     \\ 
 12236.81830 &  -17.61 &    1.44  &              K                     \\ 
 12515.95280 &  -25.82 &    1.44  &              K                     \\ 
 12574.90240 &  -34.38 &    1.45  &              K                     \\ 
 12836.11950 &  -35.49 &    1.45  &              K                     \\ 
 12989.72665 &  -42.44 &    1.35  &              K                     \\ 
 13240.00935 &  -36.93 &    1.11  &              K                     \\ 
 13961.11345 &  -24.51 &    0.64  &              K                     \\ 
 14083.90911 &  -24.85 &    1.11  &              K                     \\ 
 14397.92588 &  -10.40 &    1.05  &              K                     \\ 
 14719.07441 &   -6.08 &    1.09  &              K                     \\ 
 14809.83689 &   -0.23 &    1.14  &              K                     \\ 
 14838.92165 &    1.05 &    1.08  &              K                     \\ 
 14864.76650 &   -1.77 &    1.09  &              K                     \\ 
 15015.10431 &    5.07 &    1.12  &              K                     \\ 
 15016.10240 &    3.72 &    1.11  &              K                     \\ 
 15017.10744 &    3.23 &    1.19  &              K                     \\ 
 15042.12880 &    8.88 &    1.19  &              K                     \\ 
 15044.13700 &    8.69 &    1.14  &              K                     \\ 
 15049.04212 &    9.74 &    1.17  &              K                     \\ 
 15077.13808 &    8.48 &    1.05  &              K                     \\ 
 15085.13348 &    5.48 &    1.02  &              K                     \\ 
 15109.99945 &    9.62 &    1.38  &              K                     \\ 
 15133.95126 &    3.42 &    1.25  &              K                     \\ 
 15134.91132 &    6.11 &    1.22  &              K                     \\ 
 15171.97729 &    9.50 &    1.59  &              K                     \\ 
 15188.89679 &    8.46 &    1.18  &              K                     \\ 
 15231.78501 &    6.06 &    1.12  &              K                     \\ 
\enddata
\end{deluxetable}

\begin{deluxetable}{cccc}
\tabletypesize{\footnotesize}
\tablecaption{Radial Velocities of Gl\,179}
\tablenum{8}
\label{tab:rv_hip22627}
\tablewidth{0pt}
\tablehead{ 
    \colhead{JD -- 2440000}               & \colhead{Radial Velocity}  & \colhead{Uncertainty} & \colhead{Telescope}  \\
    \colhead{}  & \colhead{(\ms)}            & \colhead{(\ms)}       & \colhead{}
}
\startdata
 11580.83131 &   38.12 &    4.05  &              K                     \\ 
 11882.88790 &   37.65 &    4.87  &              K                     \\ 
 11901.00250 &   30.81 &    5.13  &              K                     \\ 
 12235.84867 &   54.65 &    6.14  &              K                     \\ 
 12536.08795 &    4.11 &    4.50  &              K                     \\ 
 12572.99093 &   28.53 &    5.14  &              K                     \\ 
 12573.95028 &   17.59 &    4.52  &              K                     \\ 
 12575.04686 &    5.25 &    4.48  &              K                     \\ 
 12575.99081 &    9.58 &    5.42  &              K                     \\ 
 12898.11579 &   -8.75 &    4.24  &              K                     \\ 
 13014.81847 &   -7.72 &    4.53  &              K                     \\ 
 13015.83165 &  -16.60 &    4.26  &              K                     \\ 
 13016.83228 &  -29.20 &    3.72  &              K                     \\ 
 13297.85507 &   -1.09 &   10.33  &              H                     \\ 
 13299.84862 &   14.31 &    5.54  &              H                     \\ 
 13302.97472 &    2.28 &    1.93  &              K                     \\ 
 13314.80754 &    8.21 &    4.36  &              H                     \\ 
 13340.72400 &    2.54 &    5.09  &              H                     \\ 
 13631.92988 &   10.84 &   15.59  &              H                     \\ 
 13644.90017 &   25.11 &   12.87  &              H                     \\ 
 13686.92171 &    9.61 &    6.64  &              H                     \\ 
 13719.83142 &   28.29 &   12.94  &              H                     \\ 
 13984.08891 &   32.57 &    2.23  &              K                     \\ 
 14130.85314 &   35.38 &    2.51  &              K                     \\ 
 14397.93848 &   39.52 &    2.15  &              K                     \\ 
 14411.93712 &   17.96 &   14.13  &              H                     \\ 
 14771.80303 &   22.44 &    4.48  &              H                     \\ 
 14778.99120 &   21.36 &    2.43  &              K                     \\ 
 14790.99549 &   13.51 &    2.39  &              K                     \\ 
 14807.91708 &   12.78 &    2.17  &              K                     \\ 
 14838.99126 &   19.51 &    1.76  &              K                     \\ 
 14846.95665 &   11.56 &    2.70  &              K                     \\ 
 14864.95689 &   14.55 &    2.42  &              K                     \\ 
 14928.73249 &    6.67 &    2.03  &              K                     \\ 
 14929.72575 &    4.21 &    2.64  &              K                     \\ 
 14934.73136 &   -8.63 &    2.88  &              K                     \\ 
 15077.10989 &    3.80 &    1.99  &              K                     \\ 
 15084.94943 &   -1.33 &    7.22  &              H                     \\ 
 15170.78663 &   -5.85 &    1.68  &              K                     \\ 
 15174.09252 &  -14.84 &    2.27  &              K                     \\ 
 15175.70740 &  -15.10 &    8.34  &              H                     \\ 
 15180.83537 &  -32.31 &    5.00  &              H                     \\ 
 15187.83709 &  -19.44 &    2.46  &              K                     \\ 
 15206.77569 &   -3.89 &    4.39  &              H                     \\ 
\enddata
\end{deluxetable}

\enddocument